\documentclass[rnote]{aa} 

\usepackage{graphicx}
\usepackage{txfonts}
\usepackage{natbib,twoopt}
\bibpunct{(}{)}{;}{a}{}{,}             

\begin{document} 

   \title{
HARPS spectropolarimetry of three sharp-lined Herbig Ae stars: 
New insights\thanks{Based on data obtained from the ESO Science Archive 
Facility. Observations were made with ESO telescopes on La Silla Paranal 
Observatory under programme IDs 085.D-0296(A), 187.D-0917(A, D), and 
089.D-0383(A).}}
\titlerunning{Insights to sharp-lined Herbig Ae stars}

   \author{
S.~P.~J\"arvinen\inst{1}
\and T.~A.~Carroll\inst{1}
\and S.~Hubrig\inst{1}
\and M.~Sch\"oller\inst{2}
\and I.~Ilyin\inst{1}
\and H.~Korhonen\inst{3,4}
\and M.~Pogodin\inst{5}
\and N.~A.~Drake\inst{6,7}
}

\authorrunning{S.~P.~J\"arvinen et al.}

\institute{Leibniz-Institut f\"ur Astrophysik Potsdam (AIP), 
An der Sternwarte 16, 14482 Potsdam, Germany\\
              \email{sjarvinen@aip.de}
\and
European Southern Observatory, 
Karl-Schwarzschild-Str.~2, 85748~Garching, Germany
\and
Finnish Centre for Astronomy with ESO (FINCA), University of Turku, 
V\"ais\"al\"antie~20, FI-21500 Piikki\"o, Finland
\and
Niels Bohr Institute \& Centre for Star and Planet Formation, 
University of Copenhagen, {\O}ster Voldgade 5, 1350 Copenhagen, Denmark 
\and
Pulkovo Observatory, Saint-Petersburg, 196140, Russia
\and
Saint Petersburg State University, 
Universitetski pr. 28, 198504 Saint Petersburg, Russia
\and
Observat\'orio Nacional/MCTI, 
Rua General Jos\'e Cristino 77, CEP 20921-400, Rio de Janeiro, RJ, Brazil
}
\date{Received; accepted}

\abstract
{}
   {
Recently, several arguments have been presented that favour a scenario in 
which the low detection rate of magnetic fields in Herbig Ae stars can be 
explained by the weakness of these fields and rather large measurement 
uncertainties. Spectropolarimetric studies involving sharp-lined Herbig Ae 
stars appear to be a promising approach for the detection of such weak 
magnetic fields. These studies offer a clear spectrum interpretation with 
respect to the effects of blending, local velocity fields, and chemical 
abundances, and allow us to identify a proper sample of spectral lines 
appropriate for  magnetic field determination.
}
   {
High-resolution spectropolarimetric observations of the three sharp-lined 
($v \sin i < 15$\,km\,s$^{-1}$) Herbig Ae stars HD\,101412, HD\,104237, and 
HD\,190073  have been obtained in recent years with the HARPS spectrograph in 
polarimetric mode. We used these archival observations to investigate the 
behaviour of their longitudinal magnetic fields. To carry out the magnetic 
field measurements, we used the multi-line singular value decomposition (SVD) 
method for Stokes profile reconstruction. 
}
   {
We carried out a high-resolution spectropolarimetric analysis of the Herbig~Ae 
star HD\,101412 for the first time. We discovered that different line 
lists yield differences in both the shape of the Stokes $V$ signatures and 
their field strengths. They could be interpreted in the context of the impact 
of the circumstellar matter and elemental abundance inhomogeneities on the 
measurements of the magnetic field. On the other hand, due to the small size 
of the Zeeman features on the first three epochs and the lack of near-IR 
observations, circumstellar and photospheric contributions cannot be estimated 
unambiguously. In the SVD Stokes~$V$ spectrum of the SB2 system HD\,104237, we 
detect that the secondary component, which is a T\,Tauri star, possesses a 
rather strong magnetic field $\left<B_{\rm z}\right>=129\pm12$\,G, while no 
significant field is present in the primary component. Our measurements of 
HD\,190073 confirm the presence of a variable magnetic field and indicate that 
the circumstellar environment may have a significant impact on the observed 
polarization features.
}
   {}


   \maketitle

%

\section{Introduction}
\label{sect:intro}

Magnetic fields are of fundamental importance for  intermediate-mass star 
formation and accretion-ejection processes
(e.g.\ \citealt{Lietal2009}). 
However, as of today, no consistent scenario exists that explains how magnetic 
fields in Herbig Ae/Be stars are generated and how they interact with the 
circumstellar environment. Studies of the magnetic field topology using 
high-resolution spectropolarimetry are extremely important. They enable us to 
improve our understanding of magnetically driven accretion and outflows in 
these stars. Recently, 
\citet{Hubrig2015} 
presented several arguments favouring a scenario in which the low detection 
rate of magnetic fields in Herbig Ae stars can be explained by the weakness of 
these fields and rather low measurement accuracies. Notably, the authors show 
that using high-resolution HARPS (High Accuracy Radial velocity Planet 
Searcher) spectropolarimetric observations, even very weak magnetic fields, of 
the order of tens of gauss, can be detected in sharp-lined 
($v \sin i < 15$\,km\,s$^{-1}$) Herbig Ae stars with an uncertainty of only a 
few gauss if the singular value decomposition (SVD) method is applied 
\citep{Hubrig2015}. 
In that work, a mean longitudinal magnetic field 
$\left<B_{\rm z}\right>=33\pm5$\,G was detected in the HARPS spectra of the 
sharp-lined Herbig Ae star PDS\,2 with $v\,\sin\,i=12\pm2$\,km\,s$^{-1}$. 
Similar accuracies have been reached, for example, for cool stars 
(e.g.\ \citealt{cool,ttauridonati}) 
with other high-resolution spectropolarimeters, like ESPaDOnS and Narval.

\begin{figure}
\centering
\includegraphics[width=.5\textwidth]{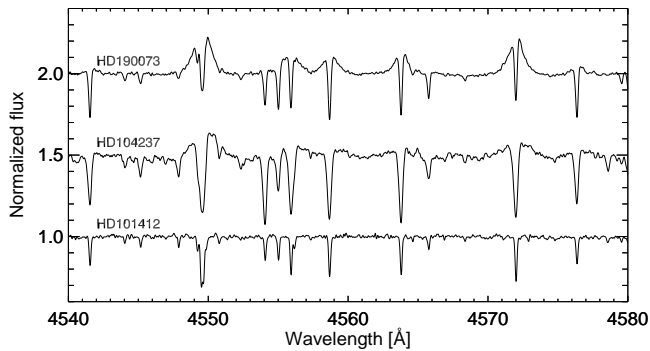}
\caption{Example spectra of the three sharp-lined Herbig Ae stars HD\,101412, 
HD\,104237, and HD\,190073.}
\label{fig:spectrum}
\end{figure}

A number of spectropolarimetric HARPS observations of three other sharp-lined 
Herbig Ae stars, \object{HD\,101412}, \object{HD\,104237}, and 
\object{HD\,190073} (see Fig.~\ref{fig:spectrum}), are publicly available in 
the European Southern Observatory (ESO) archive. In contrast to low-resolution 
spectropolarimetry where hydrogen Balmer lines are the main contributors to 
the magnetic field measurements, high-resolution polarimetric spectra 
allow us to study in detail surface abundance inhomogeneities and, in 
particular, how different elements are distributed with respect to the 
magnetic field geometry. An inhomogeneous chemical abundance distribution is 
observed most frequently on the surface of upper-main-sequence Ap/Bp stars 
with large-scale organized magnetic fields. The abundance distribution of 
certain elements in these stars is usually non-uniform and non-symmetric with 
respect to the rotation axis, but shows a kind of relationship between the 
magnetic field geometry and the element spots. As an example, rare earth spots 
are usually found in the vicinity of magnetic poles, while iron peak elements 
are concentrated closer to the magnetic equator. It is possible that a similar 
kind of relationship exists in Herbig Ae stars. In addition to the three stars 
studied in this paper and \object{PDS\,2}, the sample of 
\citet{alecian2013} 
adds only four more targets to the sharp-lined sample.

In Sect.~\ref{sect:obs}, we describe the observations and data reduction of 
these three objects, and in Sect.~\ref{sect:mf_meas} we present the method and 
results of our magnetic field measurements. Finally, in Sect.~\ref{sect:disc} 
we discuss the importance of the obtained results for our knowledge of the 
role of magnetic fields in intermediate-mass, pre-main-sequence stars.


\section{Observations and data reduction}
\label{sect:obs}

\begin{table}
\centering
\caption{Logbook of the HARPS spectropolarimetric observations.}
\label{tab:log_meas}
\begin{tabular}{cccrc}
\hline\hline
Object name & HJD & Phase & SNR & Exp.\ time\\ 
& 2450000+ &  &  & [s]\\
\hline
HD\,101412 & 6337.801 & 0.131 & 101 & 4$\times$700\\
           & 6337.835 & 0.132 & 103 & 4$\times$700\\
           & 6338.832 & 0.156 & 104 & 4$\times$700\\
           & 6339.709 & 0.176 &  92 & 4$\times$600\\
           & 6341.732 & 0.225 &  76 & 4$\times$600\\
           & 6342.634 & 0.246 &  58 & 4$\times$600\\
           & 6344.782 & 0.297 &  60 & 4$\times$600\\
\hline
HD\,104237 & 5319.714 &       & 170 & 8$\times$130\\
\hline
HD\,190073 & 5705.933 &       & 157 & 4$\times$700\\
           & 6146.715 &       &  38 & 2$\times$1800\\
\hline
\end{tabular}
\tablefoot{
The number and duration of the sub-exposures is presented in the last column. 
The phases of HD\,101412 are based on the rotation period of 42.076\,d 
determined by 
\citet{Hubrig2011a}.
}
\end{table}

Seven observations of HD\,101412, one data set for the Herbig Ae SB2 system 
HD\,104237, and two data sets for the Herbig Ae star HD\,190073 were obtained 
in recent years with the HARPS polarimeter (HARPSpol; 
\citealt{snik2008}) 
attached to ESO's 3.6\,m telescope (La~Silla, Chile). In 
Table~\ref{tab:log_meas}, the heliocentric Julian dates of the observations are 
presented in the second column, followed by the rotation phase (if known) and 
the achieved peak signal-to-noise ratio (SNR) in the Stokes~$I$ spectra. Only 
for  HD\,101412 the rotation period of 42.076\,d was determined in the past 
\citep{Hubrig2011a}. 
The archival data covers the rotation phases 0.131--0.297. The HARPSpol 
observations are usually split into several subexposures obtained with 
different orientations of the quarter-wave retarder plate relative to the beam 
splitter of the circular polarimeter. The spectra have a resolving power of 
about $R = 115\,000$ and cover the spectral range 3780--6910\,\AA{}, with a 
small gap between 5259 and 5337\,\AA{}. The reduction and calibration of the 
archive spectra was performed using the HARPS data reduction software 
available at the ESO headquarters in Germany. The normalization of the spectra 
to the continuum level consisted of several steps described in detail by 
\citet{Hubrig2013}. 
The Stokes~$I$ and $V$ parameters were derived following the ratio method 
described by 
\citet{Donati1997}. 
Null polarization spectra were calculated by combining the subexposures in 
such a way that polarization cancels out.


\section{Magnetic field measurements}
\label{sect:mf_meas}

The multi-line singular value decomposition (SVD) method was already 
successfully applied to high-resolution intensity spectra more than ten years 
ago by 
\citet{barnesSVD}, 
and the application to intermediate-resolution spectra was presented even 
earlier 
\citep{rucinski, rucinskietal}. 
The software package used to study the magnetic fields in this paper, the SVD  
method for Stokes profile reconstruction, was introduced by 
\citet{carroll2012}. 
The basic idea of SVD is similar to the principal component analysis (PCA) 
approach, where the similarity of the individual Stokes~$V$ profiles allows 
one to describe the most coherent and systematic features present in all 
spectral line profiles as a projection onto a small number of eigenprofiles. 
The line masks for each star, based on their atmospheric parameters, are 
constructed using the VALD database 
(e.g.\ \citealt{kupka2000}) 
following the basic principles stated already by 
\citet{Donati1997} 
for the least-squares deconvolution (LSD) technique. Here, we have used lines 
with intrinsic depths larger than 10\%, while hydrogen Balmer lines, strong 
\ion{He}{i}, and strong resonance lines are excluded. Also lines appearing 
close to telluric lines are removed from the mask. The SVD method is 
especially important for the extraction of polarized signals (Stokes $V$), 
which are frequently hidden below the noise level, while the reconstruction of 
the intensity profile (Stokes $I$) does not need the generally time-consuming 
application of the SVD technique.

The presence of $\delta$~Scuti-like pulsations is known 
\citep{Boehm2004} 
for the Herbig Ae star HD\,104237. These types of pulsations were detected in 
a few Herbig Ae stars in the past 
(e.g.\ \citealt{Kallinger2008}). 
Nothing is known about pulsations in the Herbig Ae stars HD\,101412 and 
HD\,190073. Since pulsations are known to have an impact on the analysis of 
the presence of a magnetic field and its strength 
(e.g.\ \citealt{schnerr2006}; \citealt{Hubrig2011b}), 
as a first step, we checked that no changes in the line profile shape or radial
velocity shifts are present in the obtained spectra on a short timescale of 
about ten minutes for HD\,101412 and on a timescale of ten to 30\,min for 
HD\,190073. These timescales correspond to the exposure times of the 
individual subexposures during the observations of these stars. For all three 
stars, the behaviour of the line profiles in the individual subexposures is 
described in Appendix~\ref{AppA}. No indication for variability on a 
short timescale was found for HD\,101412 and HD\,190073. In the following 
subsections, we present the results of the magnetic field measurements for 
each star separately. The longitudinal magnetic field 
$\left<B_{\rm z}\right>$ is measured by calculating the first-order moment of 
the Stokes $V$ profile 
(e.g.\ \citealt{Mathys1989})
\begin{equation}
\left<B_{\mathrm z}\right> = -2.4 \times 10^{11}\frac{\int \upsilon V (\upsilon)d\upsilon }{\lambda_{0}g_{0}c\int [I_{c}-I(\upsilon )]d\upsilon},
\end{equation}
where $\upsilon$ is the Doppler velocity in \,km\,s$^{-1}$, and 
$\lambda_{0}$ and $g_{0}$ are the normalization values of the wavelength and 
the average Land\'e factor.

\subsection{HD\,101412} 
\label{sect:hd101412}

Previously published mean longitudinal magnetic field measurements of the 
Herbig Ae star HD\,101412 have been carried out using low-resolution 
polarimetric spectra obtained with FORS\,1/2 (FOcal Reducer low dispersion 
Spectrograph) mounted on the 8-m Antu telescope of the VLT 
(e.g.\ \citealt{Wade2005,Wade2007}; 
\citealt{Hubrig2009,Hubrig2010,Hubrig2011a}).

\begin{figure}
\centering
\includegraphics[width=0.22\textwidth]{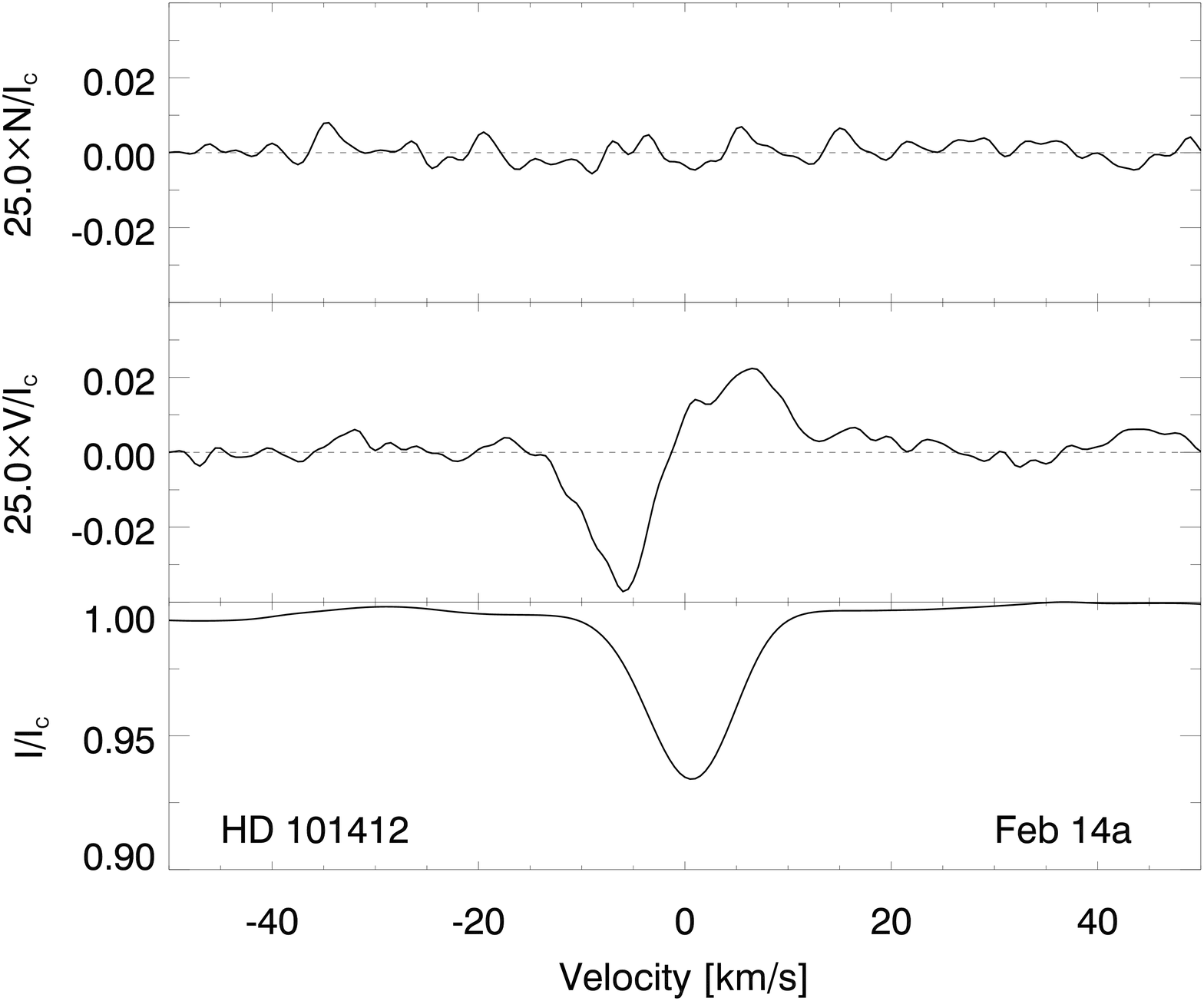}
\includegraphics[width=0.22\textwidth]{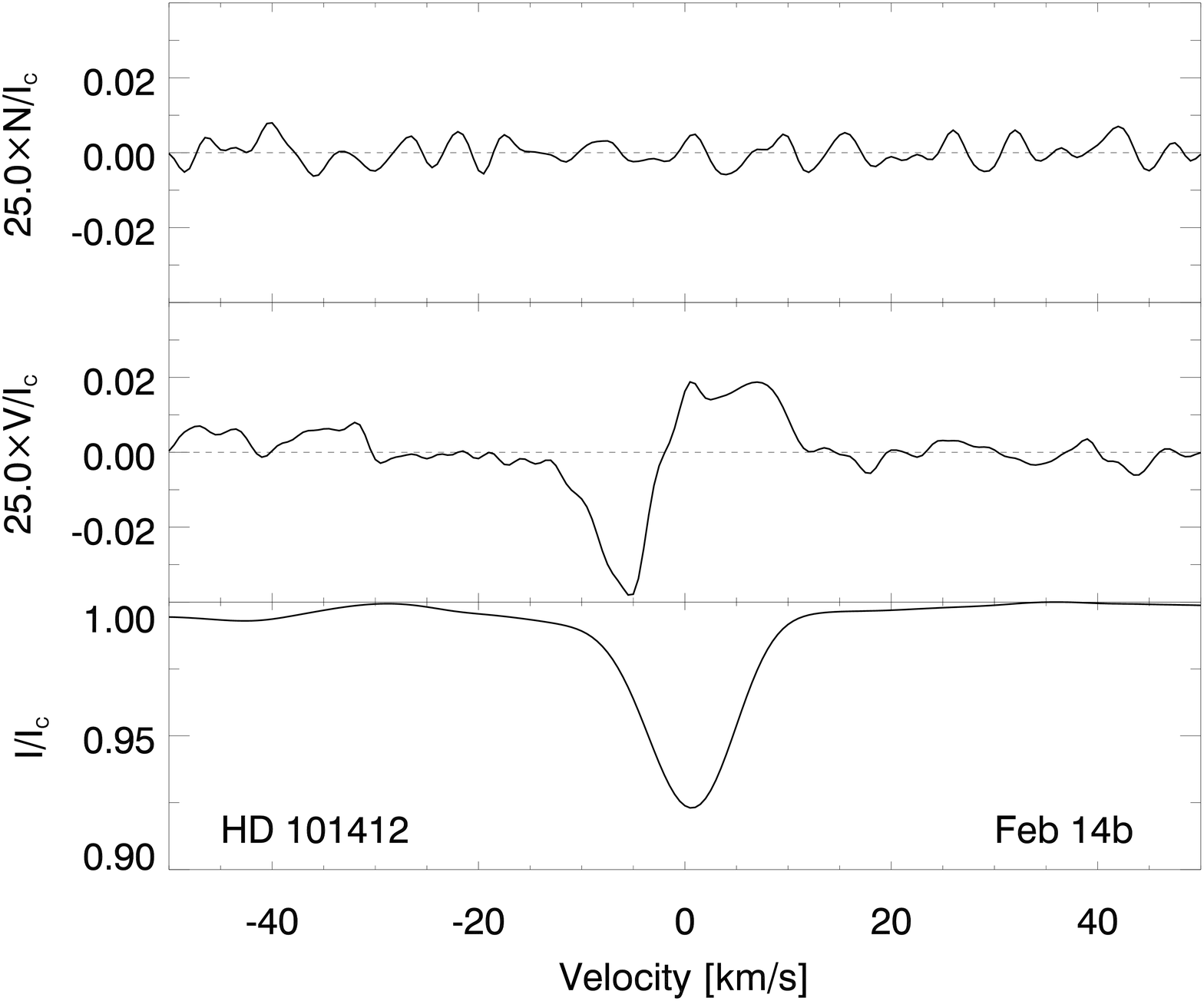}
\includegraphics[width=0.22\textwidth]{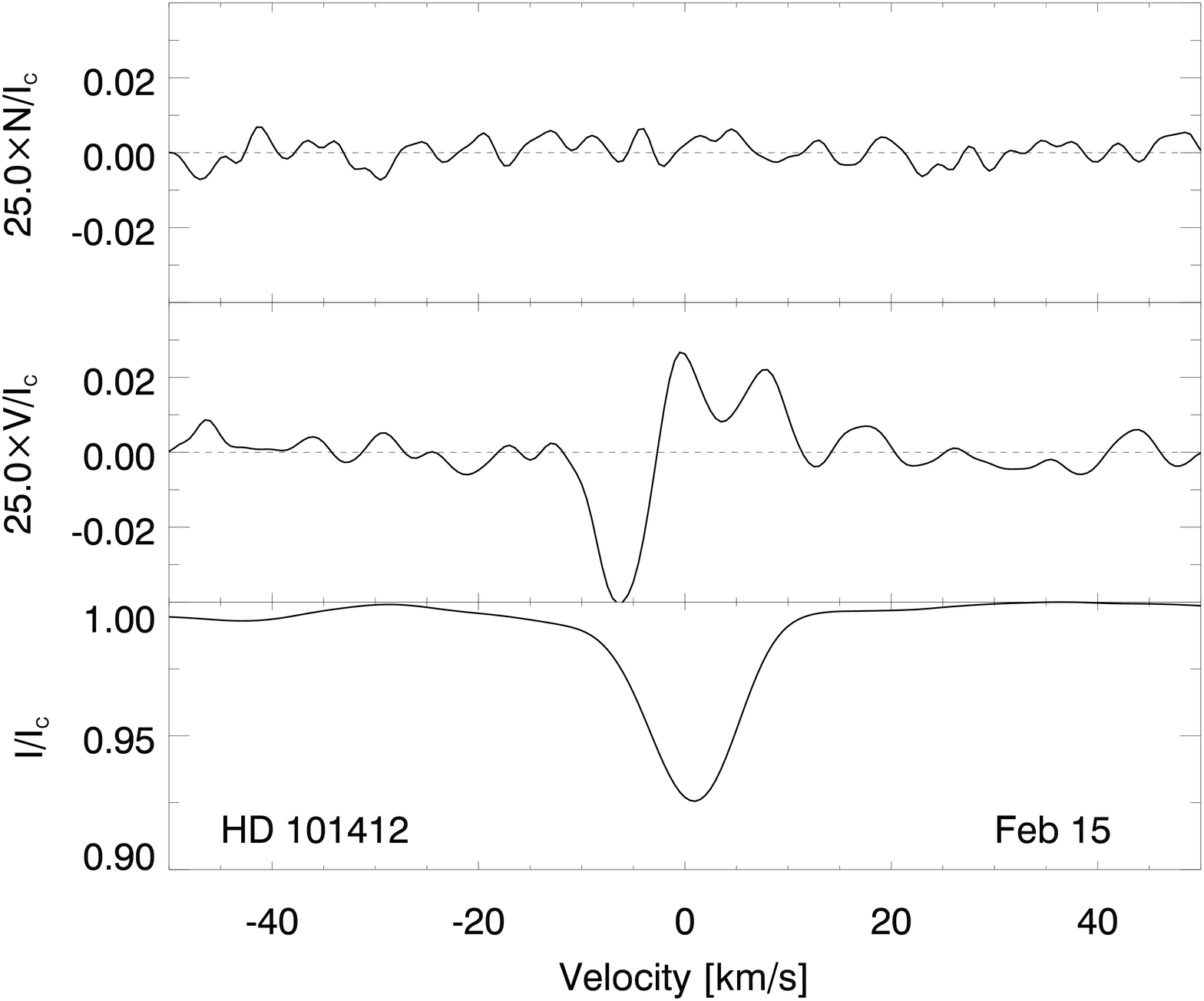}
\includegraphics[width=0.22\textwidth]{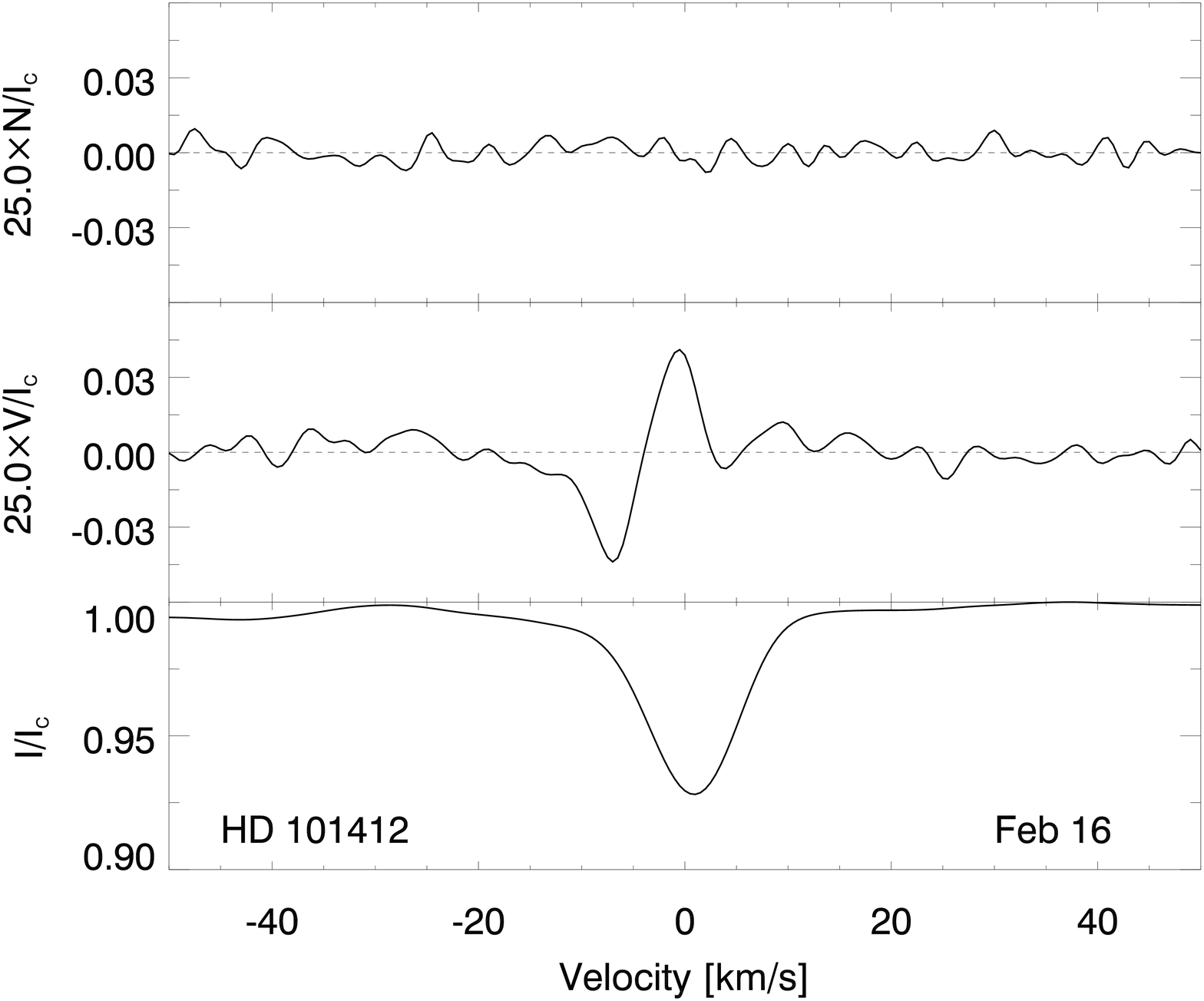}
\includegraphics[width=0.22\textwidth]{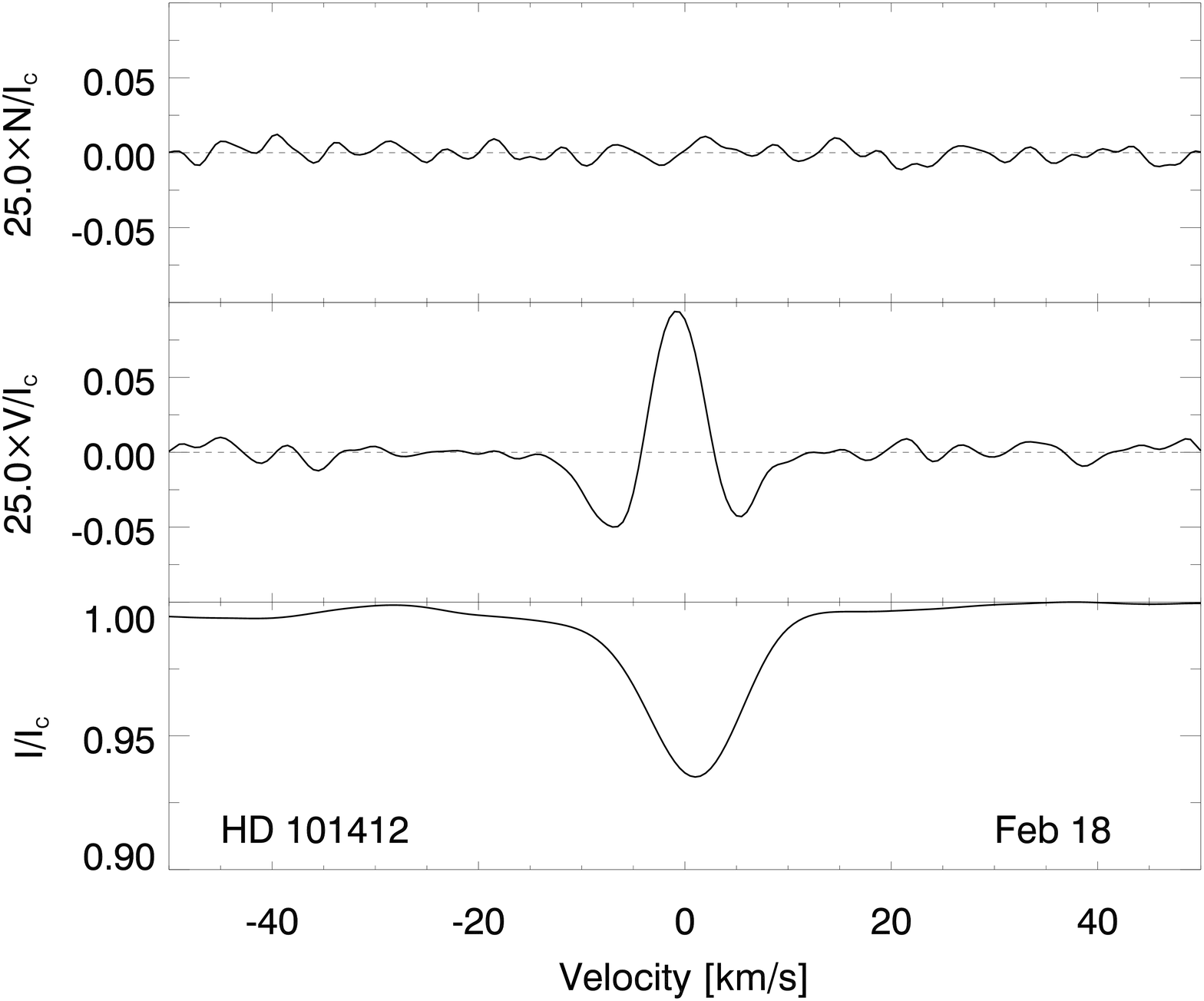}
\includegraphics[width=0.22\textwidth]{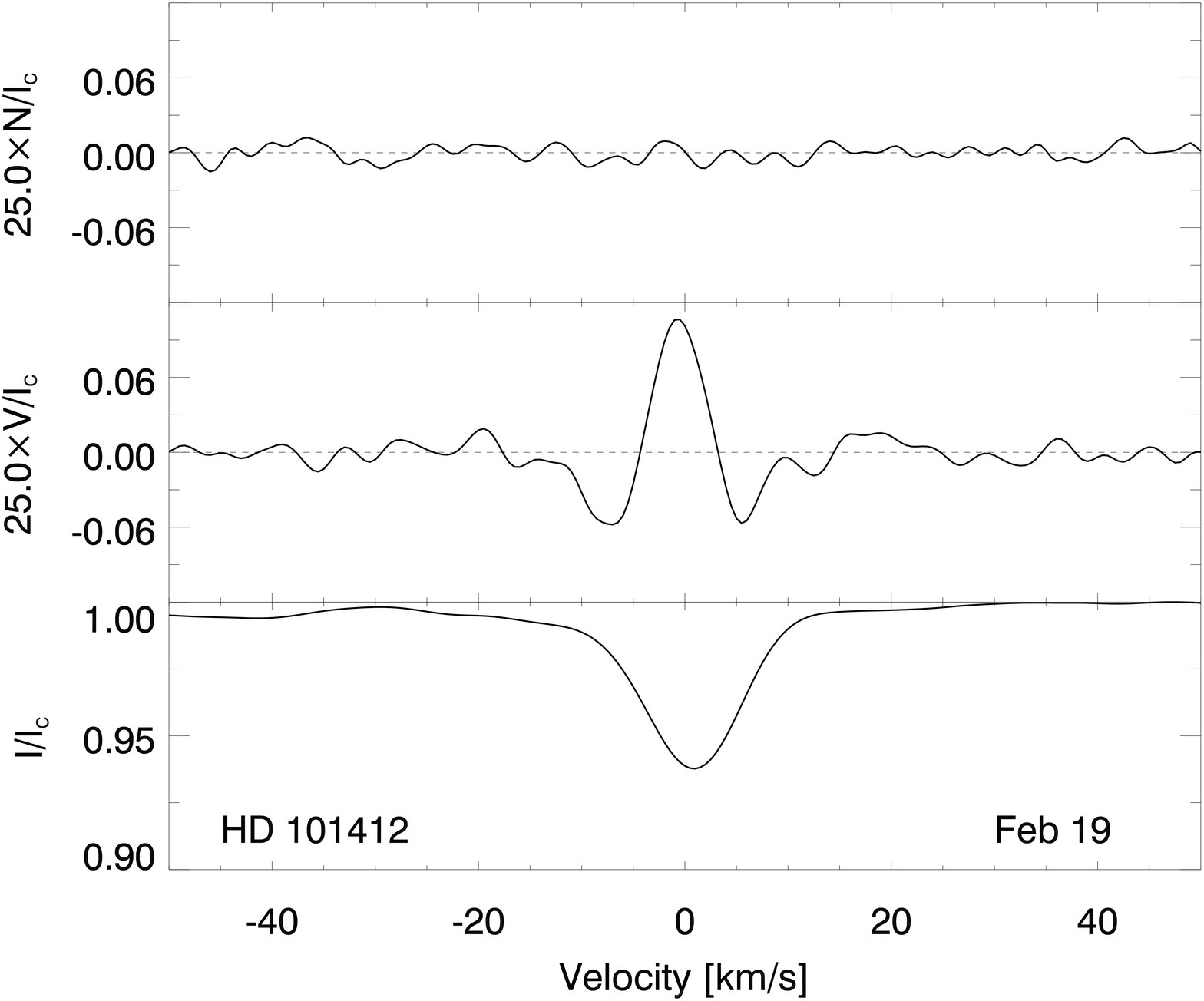}
\includegraphics[width=0.22\textwidth]{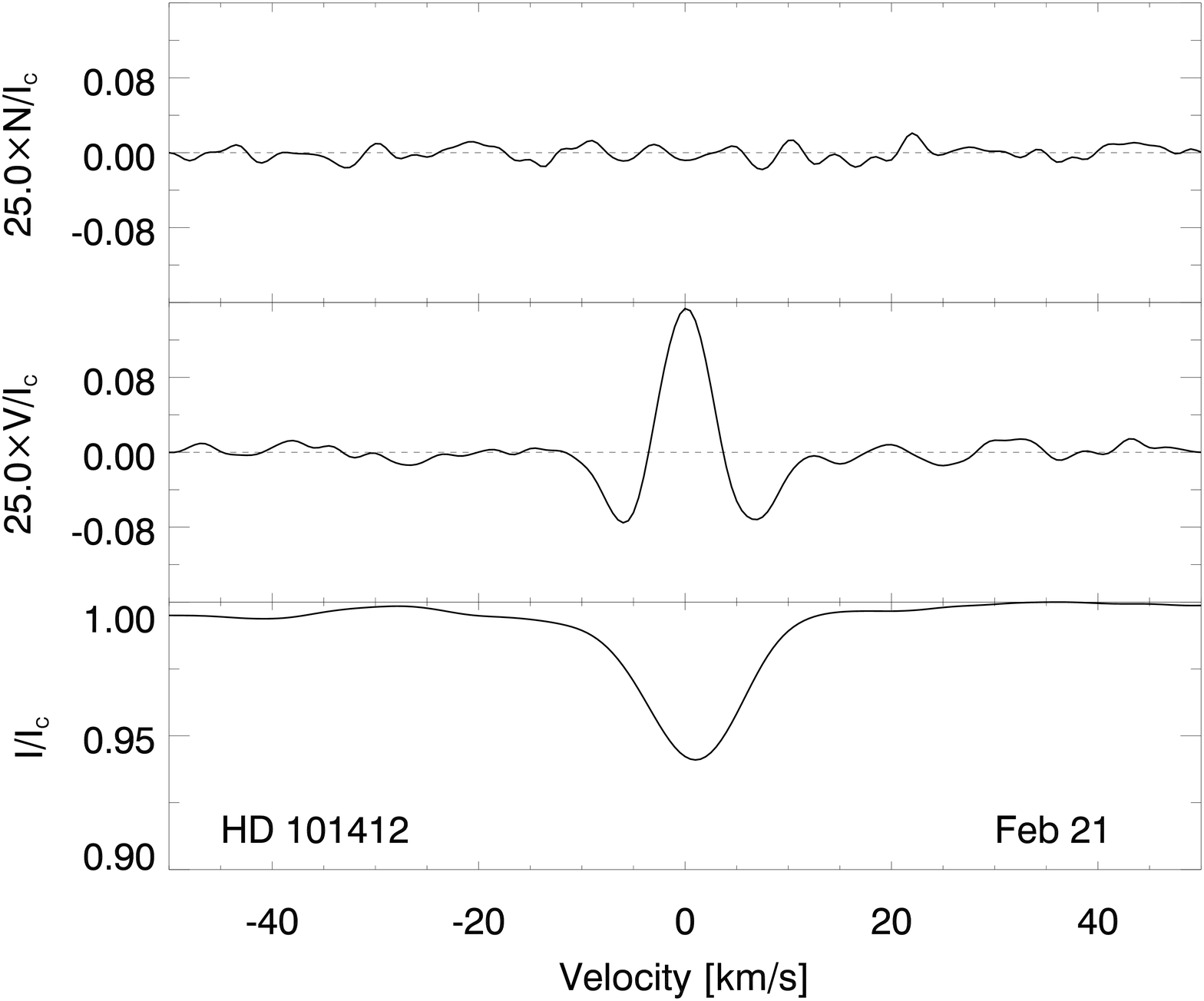}
\caption{The correspondence of SVD $I$, $V$, and $N$ profiles of 
HD\,101412 obtained using a sample of 650 lines belonging to various iron-peak 
elements assuming $T_{\rm eff}=8\,300$\,K. The profiles obtained using other 
line samples are presented in Fig.~\ref{fig:AppSVDall}.
}
\label{fig:SVDex}
\end{figure}

\begin{figure}
\centering
\includegraphics[width=0.48\textwidth]{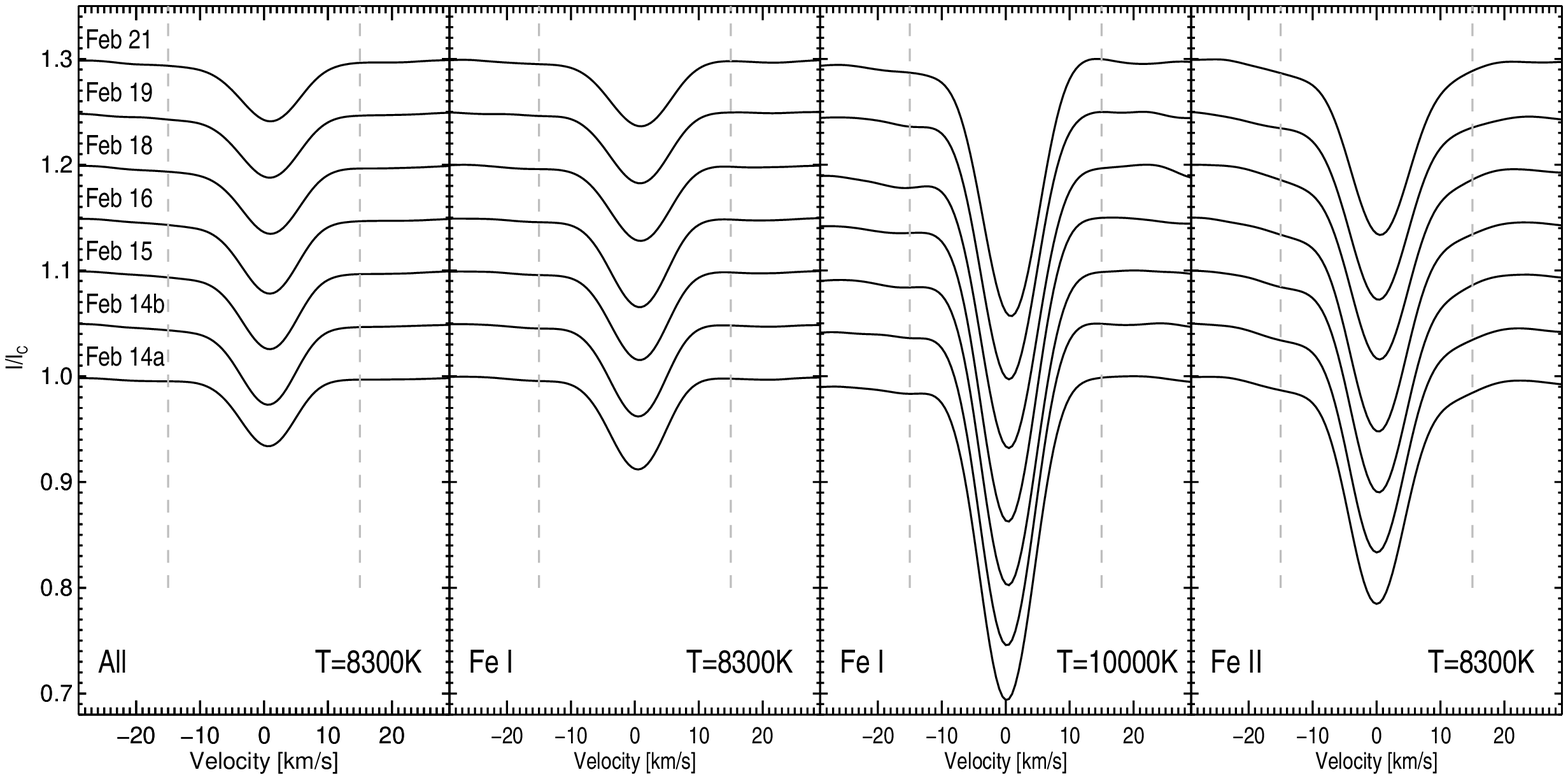}
\includegraphics[width=0.48\textwidth]{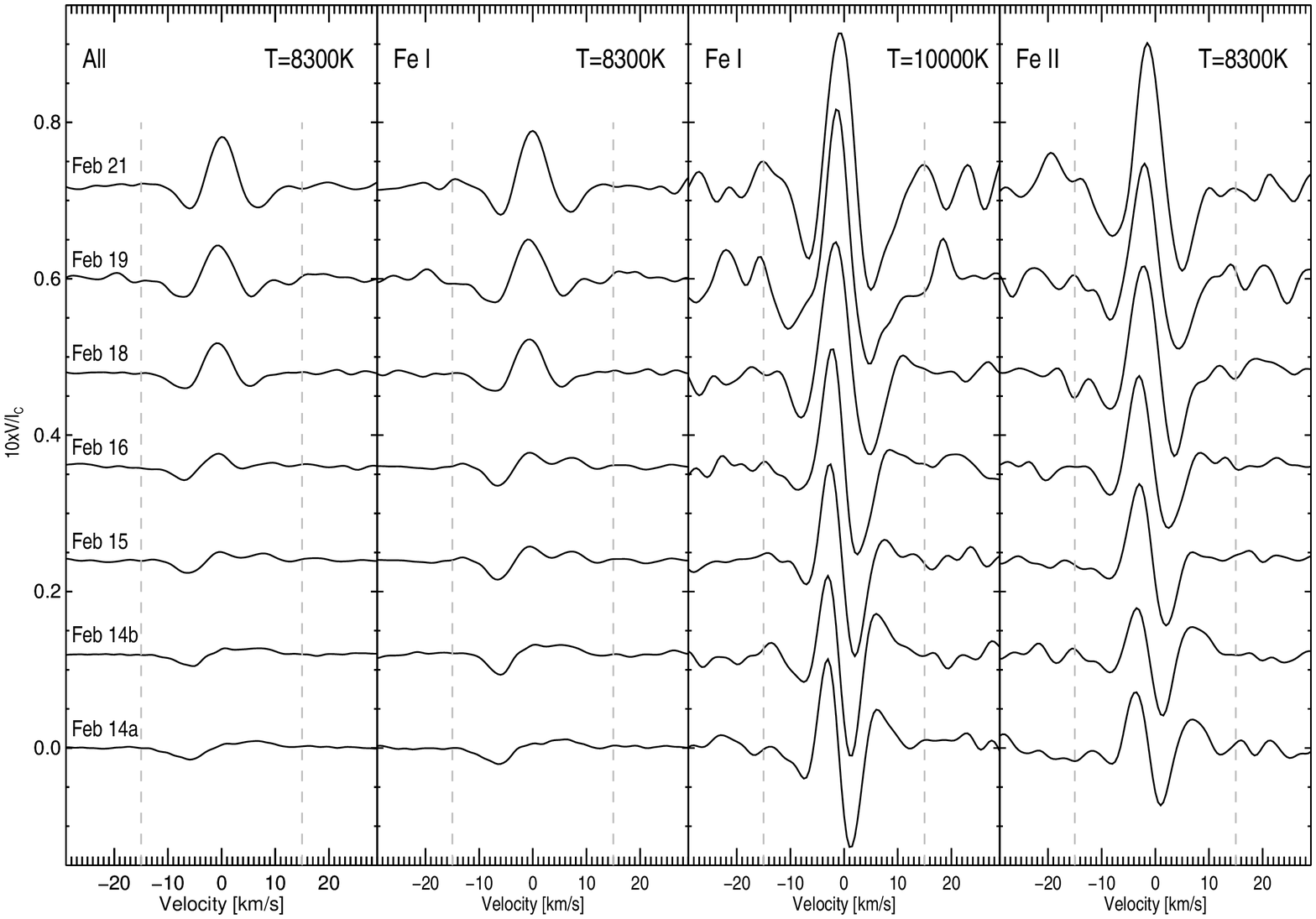}
\caption{
Time series of SVD $I$ (top) and $V$ (bottom) spectra of HD\,101412
obtained using different line masks. The profiles are shifted upwards and 
Stokes $V$ profiles are multiplied by ten for better visibility.
From left to right the results correspond to the following line samples:
the sample of 650 lines belonging to various iron-peak elements assuming 
$T_{\rm eff}=8\,300$\,K, 
the sample of 339 \ion{Fe}{i} lines ($T_{\rm eff}=8\,300$\,K), 
the sample of 29 \ion{Fe}{i} lines assuming $T_{\rm eff}=10\,000$\,K,
and the sample of 52 \ion{Fe}{ii} lines ($T_{\rm eff}=8\,300$\,K).
The vertical dashed lines indicate the velocity interval used for the magnetic 
field measurements.
}
\label{fig:SVDall}
\end{figure}

\begin{table*}
\centering
\caption{
For each sample, the signal-to-noise ratio achieved in the SVD spectra
and the measured longitudinal magnetic field are presented.
The average Land\'e factor used in the SVD measurements is also indicated.
}
\label{tab:log_all_svd}
\begin{tabular}{ccrr@{$\pm$}lrr@{$\pm$}lrr@{$\pm$}lrr@{$\pm$}l}
\hline\hline
HJD & Phase &
SNR & \multicolumn{2}{c}{$\left<B_{\rm z}\right>_{\rm All}^{\rm 8\,300K}$}     &
SNR & \multicolumn{2}{c}{$\left<B_{\rm z}\right>_{\rm Fe\,{I}}^{\rm 8\,300K}$}  &
SNR & \multicolumn{2}{c}{$\left<B_{\rm z}\right>_{\rm Fe\,{I}}^{\rm 10\,000K}$} &
SNR & \multicolumn{2}{c}{$\left<B_{\rm z}\right>_{\rm Fe\,{II}}^{\rm 8\,300K}$} \\
2450000+ & &
& \multicolumn{2}{c}{[G]} &
& \multicolumn{2}{c}{[G]} &
& \multicolumn{2}{c}{[G]} &
& \multicolumn{2}{c}{[G]} \\
\hline
 & &
\multicolumn{3}{c}{$\bar{g}_{\rm eff} = 1.29$} &
\multicolumn{3}{c}{$\bar{g}_{\rm eff} = 1.40$} &
\multicolumn{3}{c}{$\bar{g}_{\rm eff} = 1.25$} &
\multicolumn{3}{c}{$\bar{g}_{\rm eff} = 1.15$} \\
\hline
6337.801 & 0.131 & 4982 & $-$153  & 10 & 3325 & $-$140 & 11 & 873 & $-$16 & 12 & 1379 & $-$6 & 12 \\
6337.835 & 0.132 & 5253 & $-$112  & 8  & 3483 & $-$123 & 12 & 944 & 8     & 14 & 1452 & $-$3 & 12 \\
6338.832 & 0.156 & 5305 & $-$102  & 11 & 3608 & $-$110 & 8  & 959 & 29    & 16 & 1445 & 22   & 15 \\
6339.709 & 0.176 & 4472 & $-$63   & 14 & 2955 & $-$76  & 12 & 771 & 55    & 17 & 1217 & 39   & 16 \\
6341.732 & 0.225 & 3486 & $-$19   & 11 & 2306 & $-$36  & 14 & 620 & 56    & 21 & 1024 & 83   & 18 \\
6342.634 & 0.246 & 2535 & $-$13   & 19 & 1659 & $-$20  & 18 & 443 & 67    & 20 & 764  & 75   & 19 \\
6344.782 & 0.297 & 2751 & 22      & 19 & 1849 & 12     & 17 & 498 & 65    & 21 & 779  & 63   & 18 \\
\hline
\end{tabular}
\end{table*}

The fundamental parameters, $T_{\rm eff}=8\,300$\,K and $\log\,g=3.8$, and the 
projected rotation velocity, $v\,\sin\,i=3\pm2$\,km\,s$^{-1}$, were determined 
by 
\citet{cowley2010}. 
The inspection of the SVD $I$, $V$, and $N$ spectra in the phase range 
0.131--0.297 obtained using a sample of 650 metallic lines reveals that the 
amplitude of the Zeeman features increases towards the later epochs and that 
the majority of these features have a shape typical of that observed in 
classical magnetic stars during crossover, i.e.\ in the phase interval when 
the magnetic field changes its polarity 
(e.g.\ \citealt{Hubrig2014}). 
For all observations, the null spectra appear flat, indicating the absence of 
spurious polarization. Using the false alarm probability (FAP; 
\citealt{Donati1992}), 
we obtain definite magnetic field detections with FAP~$<10^{-10}$ on all 
epochs. The SVD $I$, $V$, and $N$ for this line mask are presented in 
Fig.~\ref{fig:SVDex} (see also Fig.~\ref{fig:AppSVDall}, first column). 
Surprisingly, we detect in the SVD Stokes~$V$ profiles obtained for the first 
three epochs distinct features appearing as a second maximum in the red wings 
of the Zeeman profiles. Furthermore, these profiles seem to be slightly 
shifted to the blue by about 4\,km\,s$^{-1}$. This behaviour probably 
indicates that we observe contamination by the surrounding warm circumstellar 
(CS) matter in the form of wind. This scenario seems to be in agreement with 
the previous finding of 
\citet{Hubrig2009}, 
who reported a strong contamination of UVES spectra of this star by weak lines 
of neutral and ionized iron (see their Fig. 6), where the lines of neutral 
iron are  far more numerous than those of ionized iron. 

In a circumstellar environment, such as a wind or an accretion  disk, 
lines may form over a relatively large volume, and the field topology may be 
therefore complex not only in latitude and azimuth, but in radius as well. The 
motion of the CS gas is expected to be governed by the magnetic field, 
especially in the regions of the stellar wind, which is generally assumed to 
be accelerated by a magnetic centrifuge. The presence of a clear shift of 
4\,km\,s$^{-1}$ in sharp spectral lines with $v \sin i$ values of about 
3\,km\,s$^{-1}$ probably indicates the non-photospheric origin of the detected 
features. As of today, no consistent scenario exists that 
explains how the magnetic fields in Herbig Ae stars are generated and how they 
interact with the circumstellar environment. On the other hand, it is probably 
possible to separate photospheric and CS contributions if very high SNR 
spectra in the visual and in the near-IR, recorded over the rotation cycle, 
become available in the future. Spectropolarimetric observations of 
photospheric lines are usually used to determine the geometry of the global 
magnetic field, while spectropolarimetry of accretion diagnostic lines (e.g., 
\ion{He}{i} 5876\,\AA{}, the \ion{Na}{i} doublet, the Balmer lines) probes the 
accretion topology in accretion columns, which are perturbed by interactions 
with the disk. The combination of observations in the visual with those in the 
near-IR appears most useful, since previous near-IR studies of Herbig Ae stars 
indicate that the \ion{He}{i} 1083~\,nm line is an excellent diagnostic to 
probe inflow (accretion) and outflow (winds) in the star-disk interaction 
region (\citealt{edwards1mm}).

To confirm our suspicion with respect to the impact of the CS contamination, 
we studied the behaviour of the SVD Stokes~$V$ profiles calculated with a 
second mask, containing 339 neutral iron lines forming at 
$T_{\rm eff}=8\,300$\,K and with a third mask at a significantly higher 
effective temperature, $T_{\rm eff}=10\,000$\,K containing 29 neutral iron 
lines. The results are presented in Fig.~\ref{fig:SVDall} in the second and 
the third columns (see also Fig.~\ref{fig:AppSVDall}). Noteworthy, the SVD 
Stokes~$V$ profiles calculated with a mask at a significantly higher effective 
temperature completely lack distinct features in the red wings of the Zeeman 
profiles, and show no shifts. Also our analysis of the SVD profiles obtained 
with a fourth mask containing 52 ionized iron lines presented in the fourth 
column of Fig.~\ref{fig:SVDall} and in Fig.~\ref{fig:AppSVDall} shows no 
features in the red wings of the Zeeman profiles, indicating that \ion{Fe}{ii} 
lines are predominantly formed in the stellar photosphere and not in the 
circumstellar environment. The presented results demonstrate that CS 
contamination probably plays a role in the appearance of Zeeman features and 
special care has to be taken in the interpretation of the calculated 
Stokes~$V$ profiles. On the other hand, Zeeman signatures from the first three 
epochs are much weaker compared to those in the subsequent epochs, with little 
change in the relative amplitude of the Stokes $I$ profiles. If these 
signatures are indeed circumstellar in origin, it is not completely clear yet 
whether we only see the circumstellar contribution during the crossover phases 
when the Zeeman signature has low amplitude or if this contribution is a time 
variable phenomenon that happens to be strong enough at the time of the first 
three epochs of observations and becomes gradually weaker at the subsequent 
epochs. The measurements using a velocity interval of 
$\pm15$\,km\,s$^{-1}$ are summarized in Table~\ref{tab:log_all_svd}.

\begin{figure}
\centering
\includegraphics[width=0.45\textwidth]{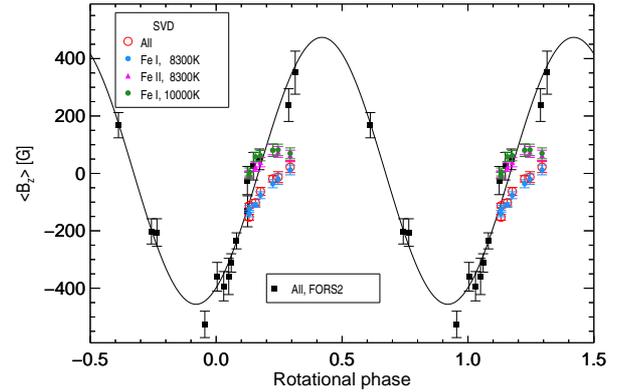}
\caption{Phase diagram for the longitudinal magnetic field measurements of 
HD\,101412 carried out using low-resolution FORS\,2 spectropolarimetric 
observations (black filled squares) and high-resolution HARPSpol observations.
Red open circles represent the SVD measurements via all metallic lines, 
blue filled circles via \ion{Fe}{i} lines at $T_{\rm eff}=8\,300$\,K, 
magenta triangles via \ion{Fe}{ii} lines at $T_{\rm eff}=8\,300$\,K, 
and green filled circles via \ion{Fe}{i} lines at $T_{\rm eff}=10\,000$\,K.
The fit in the plot refers to 
\citet{Hubrig2011a}.}
\label{fig:Bz}
\end{figure}

To visualize our results obtained for the different line masks, we present in 
Fig.~\ref{fig:Bz} the previously published FORS\,2 measurements 
\citep{Hubrig2011a} 
together with our HARPS magnetic field measurements. In this figure, we 
overplot the results for the samples of \ion{Fe}{i} lines at 
$T_{\rm eff}=8\,300$\,K and $T_{\rm eff}=10\,000$\,K, \ion{Fe}{ii} lines at
$T_{\rm eff}=8\,300$\,K, and for the line sample of all metallic lines. The 
values for the measurements via photospheric \ion{Fe}{i} and \ion{Fe}{ii} 
lines appear shifted towards more positive values of the magnetic field 
strength. This figure also clearly shows that the distribution of field values 
obtained using HARPSpol spectra is completely different compared to the 
magnetic field values determined in previous low-resolution FORS\,2 
measurements, where hydrogen Balmer lines are the main contributors to the 
magnetic field measurements. This kind of behaviour can likely be explained by 
the presence of concentration of iron-peak elements in the region of the 
magnetic equator, similar to the iron peak element spot distribution frequently 
observed in classical Ap stars 
(see e.g.\ \citealt{nonuniform}). 
The presence of iron spots was already indicated in previous studies of this 
star by 
\citet{Hubrig2011a,Hubrig2012}, 
who presented variability of equivalent width, radial velocity, line width, 
and line asymmetry using a sample of blend free iron lines.

\subsection{HD\,104237} 
\label{sect:hd104237}

The Herbig Ae star HD\,104237 was intensively studied during the last years,
in particular, because of the possible presence of a magnetic field announced 
16 years ago by 
\citet{Donati1997}. 
In their work, the authors reported on the probable detection of a weak 
magnetic field of the order of 50\,G. The star is a primary in an SB2 system 
with an orbital period of 19.86\,d 
\citep{Boehm2004}. 
A study of the chemical abundances in both components was presented by 
\citet{Cowley2013}. 
The authors assumed the following fundamental parameters of HD\,104237: 
$T_{\rm eff}=8250$\,K, $\log g=4.2$, and $v\,\sin\,i=8$\,km\,s$^{-1}$, for the 
primary component, and $T_{\rm eff}=4800$\,K, $\log g=3.7$, and 
$v\,\sin\,i=12$\,km\,s$^{-1}$, for the secondary. The primary is a 
$\delta$~Scuti-like pulsator with frequencies ranging between 28.5 and 
35.6\,d$^{-1}$ 
\citep{Boehm2004}. 
The impact of pulsations on the Stokes~$I$ profiles is presented in the 
Appendix.

As a result of the pulsation induced changes detected in the line profiles 
during the observation, the final Stokes~$V$ spectrum is expected to lead to 
the wrong value for the longitudinal magnetic field. The question of how 
pulsations affect the magnetic field measurements is not yet solved in spite 
of the fact that the number of magnetic studies of pulsating stars is 
gradually increasing. The effect of pulsations on magnetic field measurements 
was for the first time discussed by 
\citet{schnerr2006} and \citet{Hubrig2011b}. 
\citet{schnerr2006}
discussed the influence of pulsations on the analysis of the magnetic field 
strength in the $\beta$~Cephei star $\nu$~Eri in MUlti-SIte COntinuous 
Spectroscopy (MUSICOS) spectra and tried to model the signatures found in 
Stokes $V$ and $N$ spectra. Although the authors reported that they are able 
to quantitatively establish the influence of pulsations on the magnetic field 
determination through some modelling, they still detect profiles in Stokes $V$ 
and $N,$ which are the result of the combined effects of the pulsations and 
the inaccuracies in wavelength calibration that were not removed by their 
imperfect modelling of these effects. 
\citet{Hubrig2011b} 
suggested carrying out separate measurements of the line shifts between the 
spectra $(I + V)_{0}$ and $(I - V)_{0}$ in the first subexposure and the 
spectra $(I - V)_{90}$ and $(I + V)_{90}$ in the second subexposure (see
their Sect.~2 ). The final longitudinal magnetic field value is then calculated
as an average of the measurements for each subexposure. We tried to apply a 
similar procedure to the observations of HD\,104237 (see Appendix~\ref{Ap2}). 
However, the very low SNR obtained in the HARPS subexposures did not allow us 
to draw any conclusions about the strength of the magnetic field.

\begin{figure}
\centering
\includegraphics[width=0.24\textwidth]{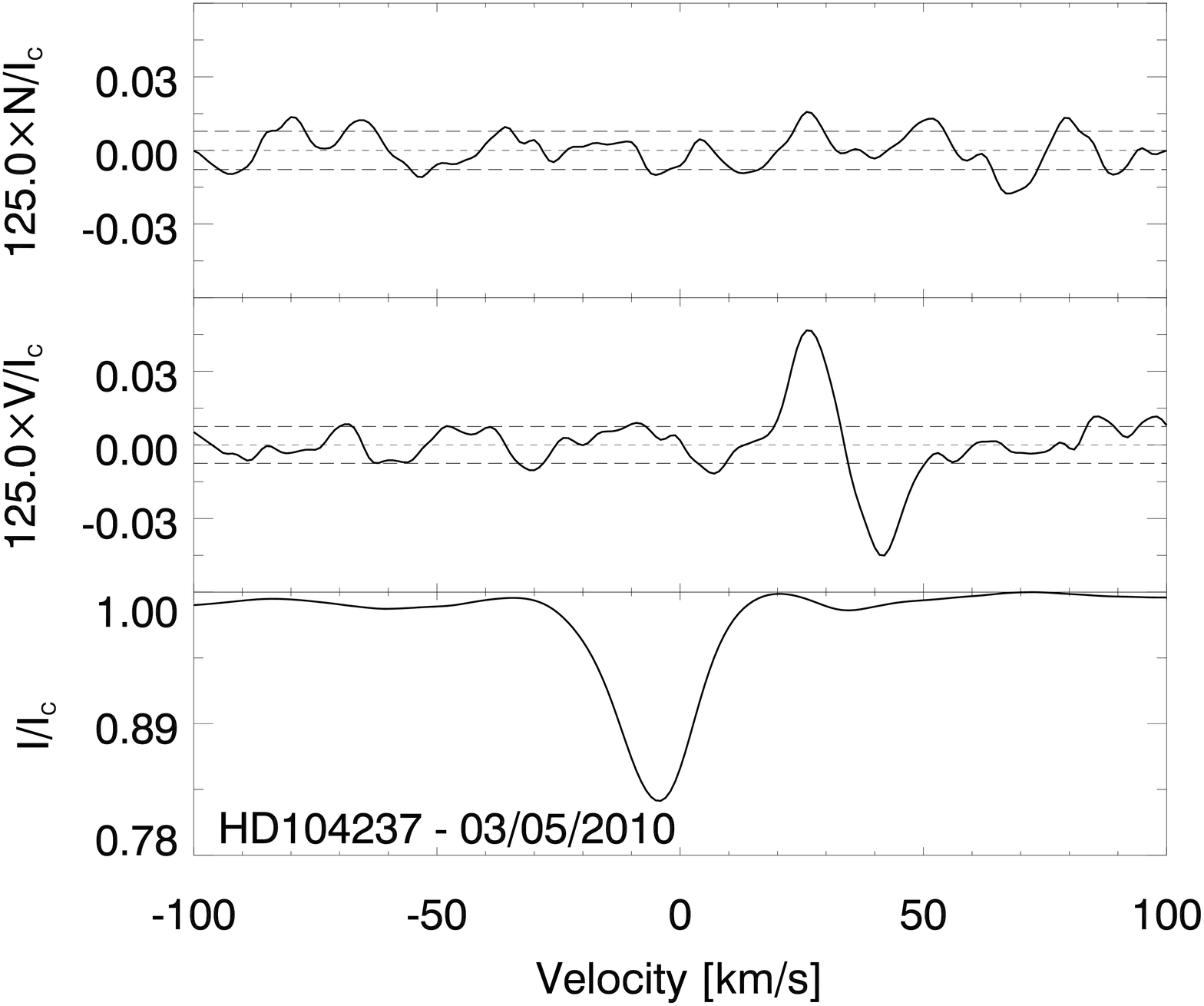}
\includegraphics[width=0.24\textwidth]{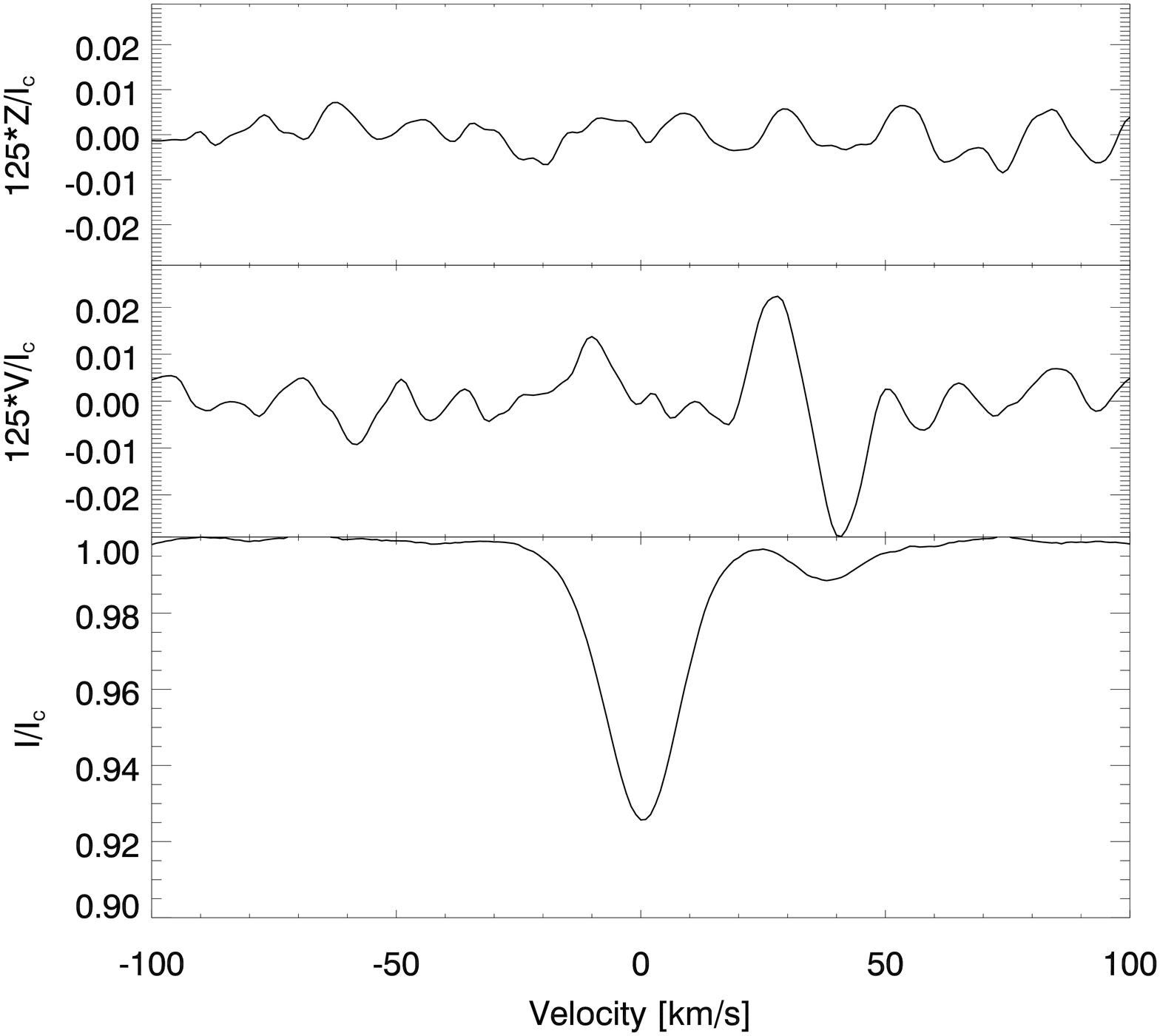}
\caption{The SVD $I$, $V$, and $N$ spectra of HD\,104237 obtained with 
different samples of metallic lines. The dashed lines in the plot on the left 
side indicate the standard deviations for the $N$ and $V$ spectra.
}
\label{fig:hd104}
\end{figure}

The results of the measurements of the mean longitudinal magnetic field 
obtained using the SVD method with 1292 metallic lines in the line mask 
are presented on the left side of Fig.~\ref{fig:hd104}. The polarization 
feature for the secondary component, which is a T\,Tauri star, stands out in 
the SVD Stokes~$V$ spectrum corresponding to 
$\left<B_{\rm z}\right> =129\pm 12$\,G with FAP~$<10^{-12}$, from the velocity 
interval 10--60\,km\,s$^{-1}$. This is the first registration of the magnetic 
field in the secondary component of the system HD\,104237, and it is in 
agreement with values derived for some other T\,Tauri stars 
\citep{TTauri}. 
If we include an additional few hundred weaker lines in the line mask, we 
detect in the SVD Stokes~$V$ spectrum of the primary a marginal polarization 
feature at the level of 13\,G (right side of Fig.~\ref{fig:hd104}). Given the 
presence of pulsations and the weakness of the detected feature, the presence 
of the magnetic field in this component remains highly uncertain.

\subsection{HD\,190073} 
\label{sect:hd190073}

The absorption and emission spectrum of the Herbig Ae star HD\,190073 was 
studied in detail by 
\citet{Catala2007} 
and more recently by 
\citet{CowleyHubrig2012}. 
The low projected rotational velocity ($v\,\sin\,i \le 9$\,km\,s$^{-1}$) may 
indicate either a very slow rotation, or a very small inclination of the 
rotation axis with respect to the line of sight. Recent interferometric 
observations in the infrared are best interpreted in terms of a circumstellar 
disk seen nearly face-on 
\citep{Eisner2004}. 
Also IUE observations indicate a low inclination angle of the rotation axis 
\citep{Hubrig2009}.

The first measurement of a longitudinal magnetic field in HD\,190073 was 
published by 
\citet{Hubrig2006} 
indicating the presence of a weak longitudinal magnetic field 
$\left<B_{\rm z}\right> =84\pm30$\,G measured on FORS\,1 low-resolution 
spectra at a 2.8$\sigma$ level. Later on, this star was studied by 
\citet{Catala2007} 
with ESPaDOnS observations, who confirmed the presence of a magnetic field, 
$\left<B_{\rm z}\right> =74\pm10$\,G, at a higher significance level. A 
longitudinal magnetic field $\left<B_{\rm z}\right> =104\pm19$\,G was reported 
by 
\citet{Hubrig2009} 
using FORS\,1 measurements. For a number of years, the magnetic field of this 
star appeared to be rather stable showing positive polarity. More recent 
observations of this star during 2011 July and 2012 October by 
\citet{Alecian2013b} 
with ESPaDOnS and Narval revealed variations of the Zeeman signature in the 
LSD spectra on timescales of days to weeks. The authors suggested that the 
detected variations of the Zeeman signatures are the result of the interaction 
between the fossil field and the ignition of a dynamo field generated in the 
newly-born convective core. Wade (priv.\ comm.) stated recently that these 
results might be due to issues with the instrument or data reduction.

\begin{figure}
\centering
\includegraphics[width=0.24\textwidth]{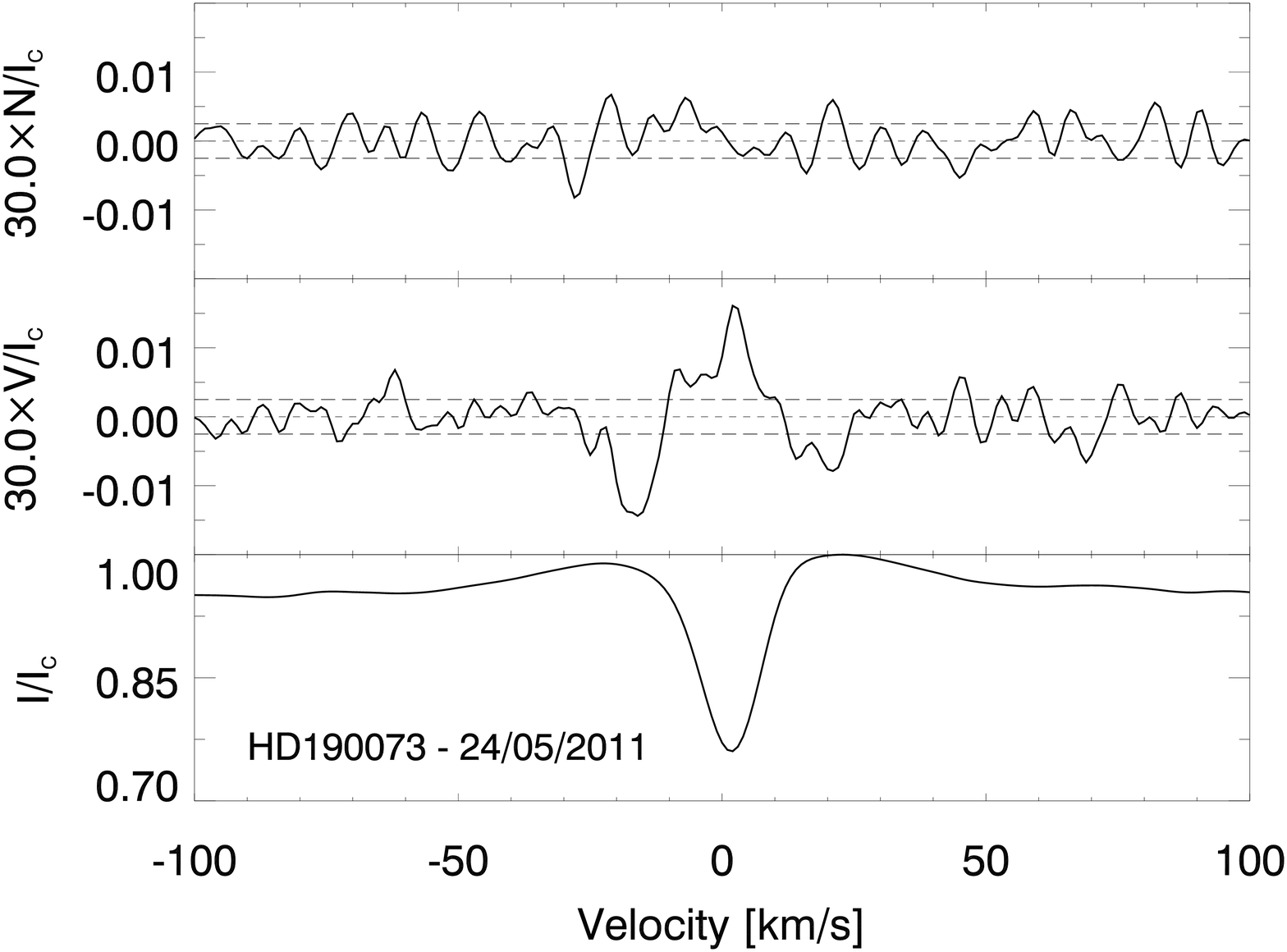}
\includegraphics[width=0.24\textwidth]{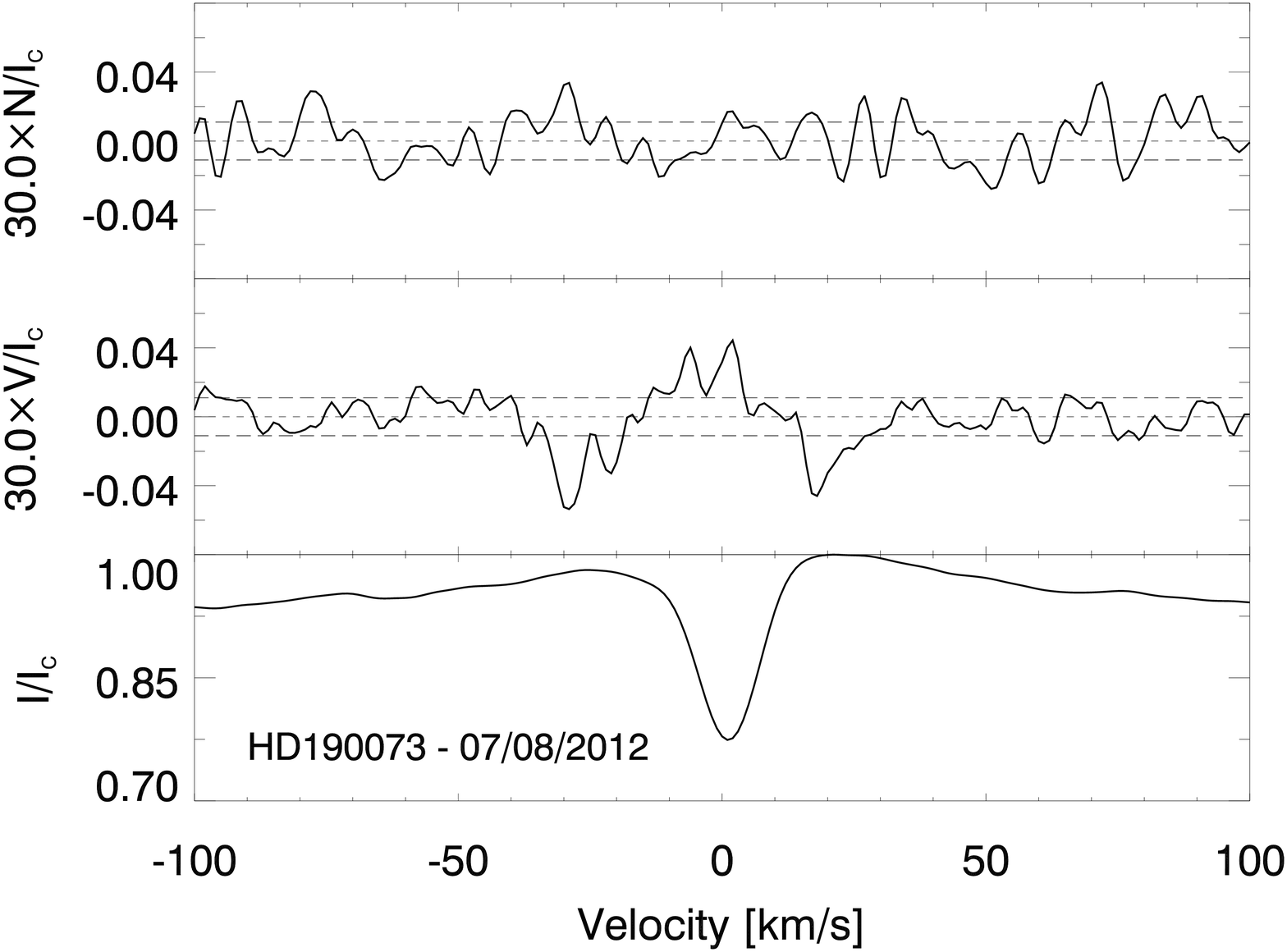}
\caption{The SVD $I$, $V$, and $N$ spectra of HD\,190073 obtained using a 
sample of 287 metallic lines at the first epoch and 522 metallic lines at 
the second epoch.The dashed lines indicate the standard deviations for the 
$N$ and $V$ spectra.
}
\label{fig:hd190}
\end{figure}

The results of our measurements of the mean longitudinal magnetic field 
obtained using the SVD method with 287 metallic lines in the line mask for 
the first epoch on 2011 May 24 are presented on the left side of 
Fig.~\ref{fig:hd190}. At this epoch, we obtain a definite detection 
$\left<B_{\rm z}\right>=-8\pm6$\,G with a false alarm probability smaller 
than $10^{-8}$. Both the shape of the profile and the field strength are in 
agreement with those of 
\citet{Alecian2013b}, 
who used the same HARPS observations as we did. For the second epoch on 2012 
August 7, we achieve a marginal detection $\left<B_{\rm z}\right>=-15\pm10$\,G 
with a false alarm probability of $8 \times 10^{-4}$. For this epoch, we had 
to employ 522 metallic lines due to very poor SNR. The resulting mean 
Stokes~$I$, Stokes~$V$, and null profiles are presented on the right side of 
Fig.~\ref{fig:hd190}. In both cases, the  velocity interval used was 
$\pm40$\,km\,s$^{-1}$. The appearance of polarization features, especially on 
the second epoch in 2012, is not easy to interpret: instead of one Zeeman 
feature in the SVD Stokes~$V$ profile, we detect two Zeeman features of 
different shapes corresponding to emission wings around the central absorption 
in the Stokes~$I$ profile (see also Fig.~\ref{fig:spectrum}). This finding 
may indicate that the CS environment has an impact on the configuration of the 
magnetic field of HD\,190073. No significant fields could be determined from 
the null spectra, indicating that any noticeable spurious polarization is 
absent.


\section{Discussion}
\label{sect:disc}

The behaviour of the magnetic field of HD\,101412 over the rotation period was 
previously based only on low-resolution polarimetric spectra. We used 
high-resolution HARPSpol spectra  to investigate the spectral variability on a 
short timescale and the behaviour of the longitudinal magnetic field over a 
part of the rotation cycle. Over the time interval covered by the HARPS 
spectra, the longitudinal magnetic field changes sign from negative to 
positive polarity. The results of our measurements based on different line 
samples indicate the presence of differences in the shape of Zeeman features 
in the Stokes~$V$ profiles and in the field strength. For the first time, they 
are discussed in the context of the impact of the circumstellar matter and 
elemental abundance inhomogeneities on the measurements of the magnetic field.

In the SVD Stokes~$V$ spectrum of HD\,104237, we detect a rather strong 
polarization feature at the position corresponding to the secondary component, 
which is a T\,Tauri star. The measured magnetic field is 
$\left<B_{\rm z}\right> =129\pm 12$\,G. We only detected a marginal 
polarization feature at the level of $\left<B_{\rm z}\right> =13\pm 11$\,G in 
the Stokes~$V$ spectrum of the primary component. Since such SB2 systems with 
sharp-lined components are extremely rare -- no other system with similar 
spectral characteristics is known to date -- it is important to carry out
a follow-up study with highly accurate HARPSpol spectra.

Our analysis of HD\,190073 confirms the presence of a variable magnetic field. 
The shape of the detected polarization features indicates a significant 
contribution of the CS environment in this star. Polarimetric spectra at 
higher SNR are urgently needed to understand the magnetic field topology.

Magnetospheric accretion has been well established for T\,Tauri stars and 
depends on a strong ordered predominantly dipolar field channelling 
circumstellar disk material to the stellar surface via accretion streams. To 
properly understand the magnetospheres of Herbig Ae stars and their 
interaction with the circumstellar environment presenting a combination of 
disk, wind, accretion, and jets, knowledge of the magnetic field strength 
and topology is indispensable. Progress in understanding the 
disk-magnetosphere interaction can, however, only come from studying a 
sufficient number of targets in detail to look for patterns encompassing 
Herbig Ae stars.

\begin{acknowledgements}
       We thank the anonymous referee for valuable comments.
       SPJ acknowledges support from 
\emph{Deut\-sche For\-schungs\-ge\-mein\-schaft, DFG\/} 
project number JA~2499/1--1.
       NAD thanks Saint Petersburg State University, Russia, for research grant 
6.38.18.2014 and FAPERJ, Rio de  Janeiro, Brazil, for Visiting Researcher grant 
E-26/200.128/2015.
\end{acknowledgements}


\bibliographystyle{aa}
\bibliography{aa26903}

\begin{thebibliography}{36}
\expandafter\ifx\csname natexlab\endcsname\relax\def\natexlab#1{#1}\fi

\bibitem[{{Alecian} {et~al.}(2013{\natexlab{a}}){Alecian}, {Neiner}, {Mathis},
  {Catala}, {Kochukhov}, \& {Landstreet}}]{Alecian2013b}
{Alecian}, E., {Neiner}, C., {Mathis}, S., {et~al.} 2013{\natexlab{a}}, \aap,
  549, L8

\bibitem[{{Alecian} {et~al.}(2013{\natexlab{b}}){Alecian}, {Wade}, {Catala},
  {Grunhut}, {Landstreet}, {Bagnulo}, {B{\"o}hm}, {Folsom}, {Marsden}, \&
  {Waite}}]{alecian2013}
{Alecian}, E., {Wade}, G.~A., {Catala}, C., {et~al.} 2013{\natexlab{b}},
  \mnras, 429, 1001

\bibitem[{{Barnes}(2004)}]{barnesSVD}
{Barnes}, J.~R. 2004, \mnras, 348, 1295

\bibitem[{{B{\"o}hm} {et~al.}(2004){B{\"o}hm}, {Catala}, {Balona}, \&
  {Carter}}]{Boehm2004}
{B{\"o}hm}, T., {Catala}, C., {Balona}, L., \& {Carter}, B. 2004, \aap, 427,
  907

\bibitem[{{Carroll} {et~al.}(2012){Carroll}, {Strassmeier}, {Rice}, \&
  {K{\"u}nstler}}]{carroll2012}
{Carroll}, T.~A., {Strassmeier}, K.~G., {Rice}, J.~B., \& {K{\"u}nstler}, A.
  2012, \aap, 548, A95

\bibitem[{{Catala} {et~al.}(2007){Catala}, {Alecian}, {Donati}, {Wade},
  {Landstreet}, {B{\"o}hm}, {Bouret}, {Bagnulo}, {Folsom}, \&
  {Silvester}}]{Catala2007}
{Catala}, C., {Alecian}, E., {Donati}, J.-F., {et~al.} 2007, \aap, 462, 293

\bibitem[{{Cowley} {et~al.}(2013){Cowley}, {Castelli}, \&
  {Hubrig}}]{Cowley2013}
{Cowley}, C.~R., {Castelli}, F., \& {Hubrig}, S. 2013, \mnras, 431, 3485

\bibitem[{{Cowley} \& {Hubrig}(2012)}]{CowleyHubrig2012}
{Cowley}, C.~R. \& {Hubrig}, S. 2012, Astr.\ Nachr., 333, 34

\bibitem[{{Cowley} {et~al.}(2010){Cowley}, {Hubrig}, {Gonz{\'a}lez}, \&
  {Savanov}}]{cowley2010}
{Cowley}, C.~R., {Hubrig}, S., {Gonz{\'a}lez}, J.~F., \& {Savanov}, I. 2010,
  \aap, 523, A65

\bibitem[{{Donati} {et~al.}(2011){Donati}, {Gregory}, {Montmerle}, {Maggio},
  {Argiroffi}, {Sacco}, {Hussain}, {Kastner}, {Alencar}, {Audard}, {Bouvier},
  {Damiani}, {G{\"u}del}, {Huenemoerder}, \& {Wade}}]{ttauridonati}
{Donati}, J.-F., {Gregory}, S.~G., {Montmerle}, T., {et~al.} 2011, \mnras, 417,
  1747

\bibitem[{{Donati} {et~al.}(1997){Donati}, {Semel}, {Carter}, {Rees}, \&
  {Collier Cameron}}]{Donati1997}
{Donati}, J.-F., {Semel}, M., {Carter}, B.~D., {Rees}, D.~E., \& {Collier
  Cameron}, A. 1997, \mnras, 291, 658

\bibitem[{{Donati} {et~al.}(1992){Donati}, {Semel}, \& {Rees}}]{Donati1992}
{Donati}, J.-F., {Semel}, M., \& {Rees}, D.~E. 1992, \aap, 265, 669

\bibitem[{{Edwards} {et~al.}(2006){Edwards}, {Fischer}, {Hillenbrand}, \&
  {Kwan}}]{edwards1mm}
{Edwards}, S., {Fischer}, W., {Hillenbrand}, L., \& {Kwan}, J. 2006, \apj, 646,
  319

\bibitem[{{Eisner} {et~al.}(2004){Eisner}, {Lane}, {Hillenbrand}, {Akeson}, \&
  {Sargent}}]{Eisner2004}
{Eisner}, J.~A., {Lane}, B.~F., {Hillenbrand}, L.~A., {Akeson}, R.~L., \&
  {Sargent}, A.~I. 2004, \apj, 613, 1049

\bibitem[{{Gregory} {et~al.}(2012){Gregory}, {Donati}, {Morin}, {Hussain},
  {Mayne}, {Hillenbrand}, \& {Jardine}}]{TTauri}
{Gregory}, S.~G., {Donati}, J.-F., {Morin}, J., {et~al.} 2012, \apj, 755, 97

\bibitem[{{Hubrig} {et~al.}(2014){Hubrig}, {Carroll}, {Gonz{\'a}lez},
  {Sch{\"o}ller}, {Ilyin}, {Saffe}, {Castelli}, {Leone}, \&
  {Giarrusso}}]{Hubrig2014}
{Hubrig}, S., {Carroll}, T.~A., {Gonz{\'a}lez}, J.~F., {et~al.} 2014, \mnras,
  440, L6

\bibitem[{{Hubrig} {et~al.}(2015){Hubrig}, {Carroll}, {Sch{\"o}ller}, \&
  {Ilyin}}]{Hubrig2015}
{Hubrig}, S., {Carroll}, T.~A., {Sch{\"o}ller}, M., \& {Ilyin}, I. 2015,
  \mnras, 449, L118

\bibitem[{{Hubrig} {et~al.}(2012){Hubrig}, {Castelli}, {Gonz{\'a}lez}, {Elkin},
  {Mathys}, {Cowley}, {Wolff}, \& {Sch{\"o}ller}}]{Hubrig2012}
{Hubrig}, S., {Castelli}, F., {Gonz{\'a}lez}, J.~F., {et~al.} 2012, \aap, 542,
  A31

\bibitem[{{Hubrig} {et~al.}(2011{\natexlab{a}}){Hubrig}, {Ilyin}, {Briquet},
  {Sch{\"o}ller}, {Gonz{\'a}lez}, {Nu{\~n}ez}, {De Cat}, \&
  {Morel}}]{Hubrig2011b}
{Hubrig}, S., {Ilyin}, I., {Briquet}, M., {et~al.} 2011{\natexlab{a}}, \aap,
  531, L20

\bibitem[{{Hubrig} {et~al.}(2013){Hubrig}, {Ilyin}, {Sch{\"o}ller}, \& {Lo
  Curto}}]{Hubrig2013}
{Hubrig}, S., {Ilyin}, I., {Sch{\"o}ller}, M., \& {Lo Curto}, G. 2013, Astr.\
  Nachr., 334, 1093

\bibitem[{{Hubrig} {et~al.}(2011{\natexlab{b}}){Hubrig}, {Mikul{\'a}{\v s}ek},
  {Gonz{\'a}lez}, {Sch{\"o}ller}, {Ilyin}, {Cur{\'e}}, {Zejda}, {Cowley},
  {Elkin}, {Pogodin}, \& {Yudin}}]{Hubrig2011a}
{Hubrig}, S., {Mikul{\'a}{\v s}ek}, Z., {Gonz{\'a}lez}, J.~F., {et~al.}
  2011{\natexlab{b}}, \aap, 525, L4

\bibitem[{{Hubrig} {et~al.}(2010){Hubrig}, {Sch{\"o}ller}, {Savanov},
  {Gonz{\'a}lez}, {Cowley}, {Sch{\"u}tz}, {Arlt}, \&
  {R{\"u}diger}}]{Hubrig2010}
{Hubrig}, S., {Sch{\"o}ller}, M., {Savanov}, I., {et~al.} 2010, Astr.\ Nachr.,
  331, 361

\bibitem[{{Hubrig} {et~al.}(2009){Hubrig}, {Stelzer}, {Sch{\"o}ller}, {Grady},
  {Sch{\"u}tz}, {Pogodin}, {Cur{\'e}}, {Hamaguchi}, \& {Yudin}}]{Hubrig2009}
{Hubrig}, S., {Stelzer}, B., {Sch{\"o}ller}, M., {et~al.} 2009, \aap, 502, 283

\bibitem[{{Hubrig} {et~al.}(2006){Hubrig}, {Yudin}, {Sch{\"o}ller}, \&
  {Pogodin}}]{Hubrig2006}
{Hubrig}, S., {Yudin}, R.~V., {Sch{\"o}ller}, M., \& {Pogodin}, M.~A. 2006,
  \aap, 446, 1089

\bibitem[{{Kallinger} {et~al.}(2008){Kallinger}, {Zwintz}, \&
  {Weiss}}]{Kallinger2008}
{Kallinger}, T., {Zwintz}, K., \& {Weiss}, W. 2008, \aap, 488, 279

\bibitem[{{Kupka} {et~al.}(2000){Kupka}, {Ryabchikova}, {Piskunov}, {Stempels},
  \& {Weiss}}]{kupka2000}
{Kupka}, F.~G., {Ryabchikova}, T.~A., {Piskunov}, N.~E., {Stempels}, H.~C., \&
  {Weiss}, W.~W. 2000, Baltic Astronomy, 9, 590

\bibitem[{{Landstreet} {et~al.}(2014){Landstreet}, {Bagnulo}, \&
  {Fossati}}]{nonuniform}
{Landstreet}, J.~D., {Bagnulo}, S., \& {Fossati}, L. 2014, \aap, 572, A113

\bibitem[{{Li} {et~al.}(2009){Li}, {Dowell}, {Goodman}, {Hildebrand}, \&
  {Novak}}]{Lietal2009}
{Li}, H.-B., {Dowell}, C.~D., {Goodman}, A., {Hildebrand}, R., \& {Novak}, G.
  2009, \apj, 704, 891

\bibitem[{{Mathys}(1989)}]{Mathys1989}
{Mathys}, G. 1989, \fcp, 13, 143

\bibitem[{{Rucinski}(1992)}]{rucinski}
{Rucinski}, S.~M. 1992, \aj, 104, 1968

\bibitem[{{Rucinski} {et~al.}(1993){Rucinski}, {Lu}, \& {Shi}}]{rucinskietal}
{Rucinski}, S.~M., {Lu}, W.-X., \& {Shi}, J. 1993, \aj, 106, 1174

\bibitem[{{Schnerr} {et~al.}(2006){Schnerr}, {Verdugo}, {Henrichs}, \&
  {Neiner}}]{schnerr2006}
{Schnerr}, R.~S., {Verdugo}, E., {Henrichs}, H.~F., \& {Neiner}, C. 2006, \aap,
  452, 969

\bibitem[{{Snik} {et~al.}(2008){Snik}, {Jeffers}, {Keller}, {Piskunov},
  {Kochukhov}, {Valenti}, \& {Johns-Krull}}]{snik2008}
{Snik}, F., {Jeffers}, S., {Keller}, C., {et~al.} 2008, in Soc.\ of
  Photo-Optical Instrumentation Engineers (SPIE) Conf.\ Ser., Vol. 7014, SPIE
  Conf.\ Ser., 70140O

\bibitem[{{Wade} {et~al.}(2007){Wade}, {Bagnulo}, {Drouin}, {Landstreet}, \&
  {Monin}}]{Wade2007}
{Wade}, G.~A., {Bagnulo}, S., {Drouin}, D., {Landstreet}, J.~D., \& {Monin}, D.
  2007, \mnras, 376, 1145

\bibitem[{{Wade} {et~al.}(2005){Wade}, {Drouin}, {Bagnulo}, {Landstreet},
  {Mason}, {Silvester}, {Alecian}, {B{\"o}hm}, {Bouret}, {Catala}, \&
  {Donati}}]{Wade2005}
{Wade}, G.~A., {Drouin}, D., {Bagnulo}, S., {et~al.} 2005, \aap, 442, L31

\bibitem[{{Waite} {et~al.}(2015){Waite}, {Marsden}, {Carter}, {Petit},
  {Donati}, {Jeffers}, \& {Boro Saikia}}]{cool}
{Waite}, I.~A., {Marsden}, S.~C., {Carter}, B.~D., {et~al.} 2015, \mnras, 449,
  8

\end{thebibliography}


\appendix

\section{The study of spectral variability on short-time scales}\label{AppA}

Given the rather low SNR of the HARPS spectra, to study the spectral 
variability on short time scales, we employed the LSD technique, allowing us 
to achieve much higher SNR in the LSD spectra. The details of this technique 
can be found in the work of 
\citet{Donati1997}. 
The line masks are constructed using the VALD database 
(e.g.\ \citealt{kupka2000}).
As mentioned in Sect.~\ref{sect:mf_meas}, the reconstruction of the 
intensity profile (Stokes $I$) does not need the generally time-consuming 
application of the SVD technique.

\subsection{HD\,101412}\label{Ap1}

\begin{figure*}
\resizebox{\hsize}{!}{\includegraphics{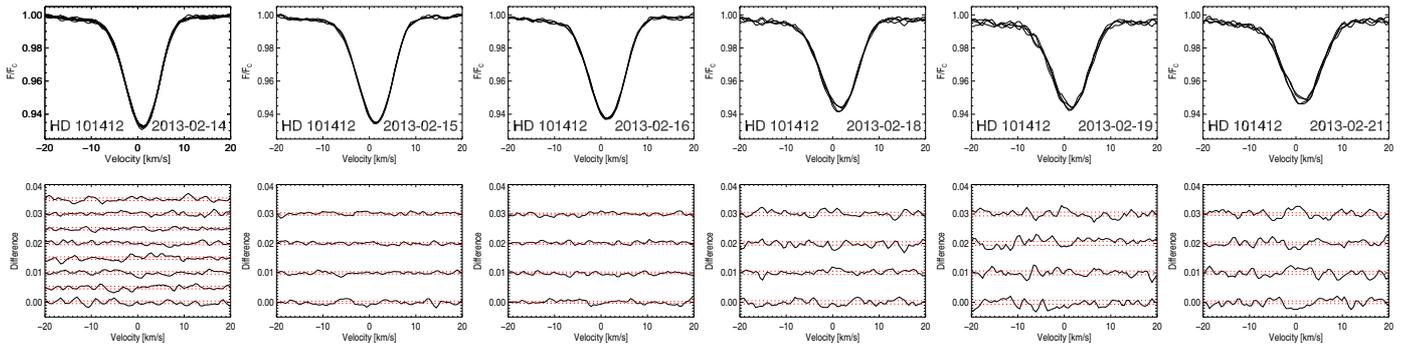}}
\caption{Top panel: comparison of the LSD Stokes~$I$ profiles of HD\,101412 
computed for individual subexposures recorded over six observing nights in 
2014 February using the line mask of 1007 metallic lines. Eight individual 
subexposures were obtained during the first night, while four subexposures were 
obtained for each following night. 
Bottom panel: differences between Stokes~$I$ profiles computed for the 
individual subexposures and the average Stokes~$I$ profile. 
The dashed lines indicate the standard deviation limits.}
\label{fig:daily}
\end{figure*}

Seven spectropolarimetric observations were obtained with the HARPS 
polarimeter on the nights from 2013 February 14 to 21. Among them, two 
individual observations have been obtained during the first night on 
February 14. Each observation was split into four subexposures with an exposure 
time of 10--12\,min, obtained with different orientations of the quarter-wave 
retarder plate relative to the polarization beam splitter of the circular 
polarimeter. In Fig.~\ref{fig:daily}, we present the comparison between the 
LSD Stokes~$I$ profiles computed for each individual subexposure recorded on 
the six different nights. The results for both observations obtained on the 
first night on February 14, each consisting of four subexposures, are 
presented in the same panel. No significant variation in the line profile or 
radial velocity is detected in the behaviour of the Stokes~$I$ profiles for 
each subexposure. Signal-to-noise values achieved in the spectra observed 
during the last three observing nights are in the range from 76 to 58, i.e.\ 
significantly  lower than those achieved in the first three epochs. For these 
last three nights, we observe very tiny changes in the cores and wings of the 
overplotted LSD Stokes~$I$ profiles with an intensity variation of the order 
of the spectral noise of about 0.3\%.

\subsection{HD\,104237}\label{Ap2}

\begin{figure}
\centering
\includegraphics[width=0.5\textwidth]{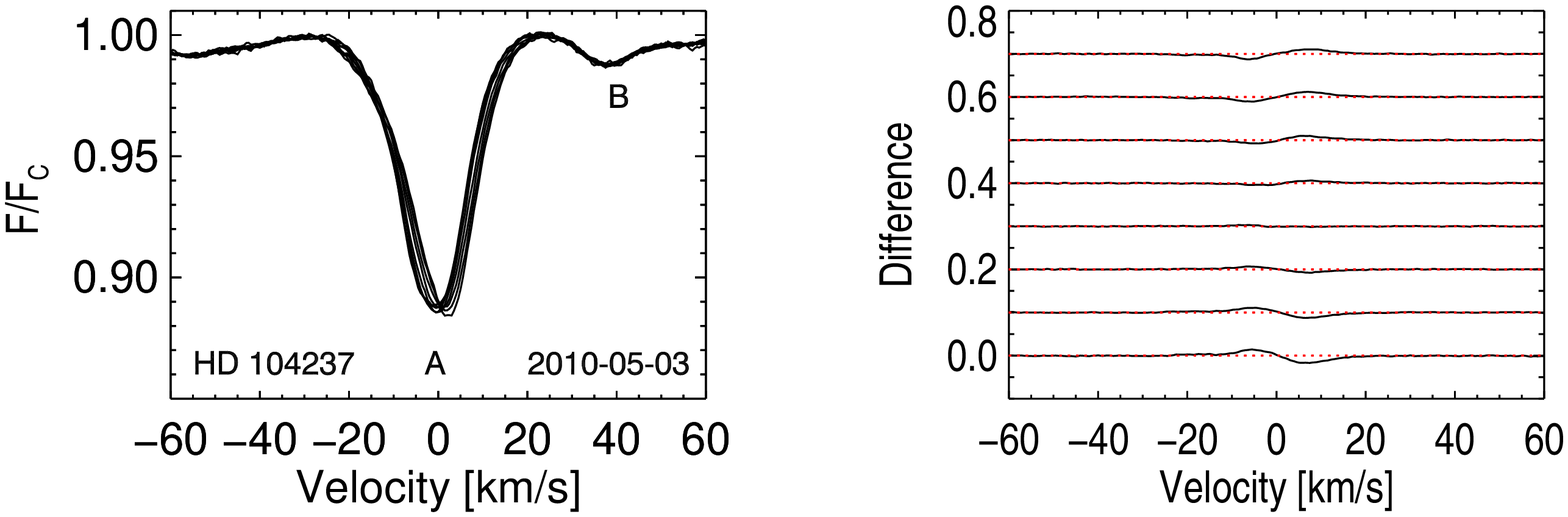}
\includegraphics[width=0.5\textwidth]{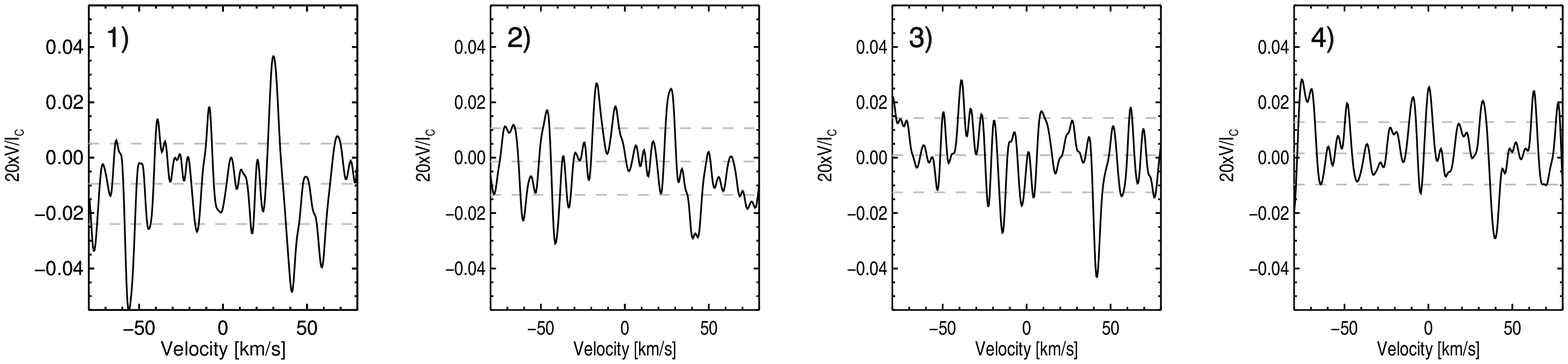}
\caption{Upper row: comparison of the LSD Stokes~$I$ profiles of HD\,104237 
computed for the individual subexposures obtained with a time lapse of 
2.2\,min (left side) and differences between Stokes~$I$ profiles computed for 
the individual subexposures and the average Stokes~$I$ profile 
with standard deviation limits indicated by the dashed lines 
(right side). Lower row: Stokes~$V$ spectra calculated for the combination of 
pairs of two subexposures with the quarter-wave plate angles separated by 90 
degrees, four pairs in all.
}
\label{fig:puls}
\end{figure}

The primary is a $\delta$~Scuti-like pulsator with frequencies ranging between 
28.5 and 35.6\,d$^{-1}$ 
\citep{Boehm2004}. 
The observation on 2010 May 3 was split into eight subexposures with an 
exposure time of 2.2\,min. The line mask included 1007 metallic lines. In 
Fig.~\ref{fig:puls} in the upper row, we present the impact of pulsations on 
the Stokes~$I$ profiles computed for each individual subexposure. Because of the 
pulsation changes in the line profiles during the observation, the final 
Stokes~$V$ spectrum is expected to lead to a wrong value of the longitudinal 
magnetic field. To prove whether the primary possesses a field, we used pairs 
of two subexposures with the quarter-wave plate angles separated by 90 
degrees, four pairs in all. However, no conclusions about the presence of a 
magnetic field in this component can be drawn due to the very low SNR 
achieved. The computed Stokes~$V$ spectra for each pair of subexposures are 
presented in the lower row of Fig.~\ref{fig:puls}.

\subsection{HD\,190073}\label{Ap3}

\begin{figure}
\centering
\includegraphics[width=0.5\textwidth]{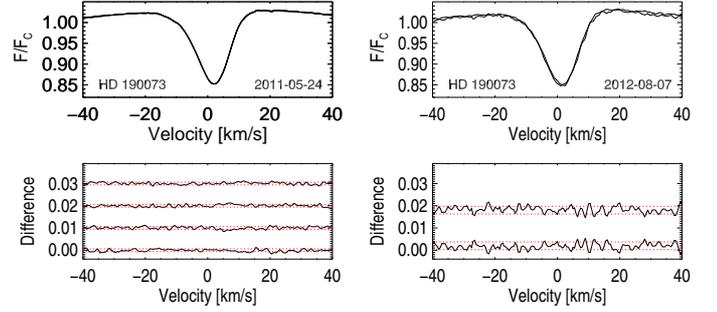}
\caption{Comparison of the LSD Stokes~$I$ profiles of HD\,190073
computed for the individual subexposures obtained on 2011 May 24 with a time 
lapse of about 10\,min (left panel) and on 2012 August 7 with a time lapse of 
30\,min (right panel). 
The dashed lines on the lower plots indicate the standard deviation 
limits.
}
\label{fig:hd190puls}
\end{figure}

This star was not reported in the literature to show $\delta$~Scuti-like 
pulsations. The line mask in our analysis included 375 metallic lines. The 
inspection of the behaviour of the LSD Stokes~$I$ profiles presented in 
Fig.~\ref{fig:hd190puls} calculated for each subexposure does not reveal any 
line profile or radial velocity variation on a timescale of about ten minutes
during the first epoch in 2011 or on a timescale of 30 minutes during the 
second epoch in 2012.

\section{The SVD profiles}\label{AppB}

\begin{figure*}
\centering
\includegraphics[width=0.20\textwidth]{Fi2g.eps}
\includegraphics[width=0.20\textwidth]{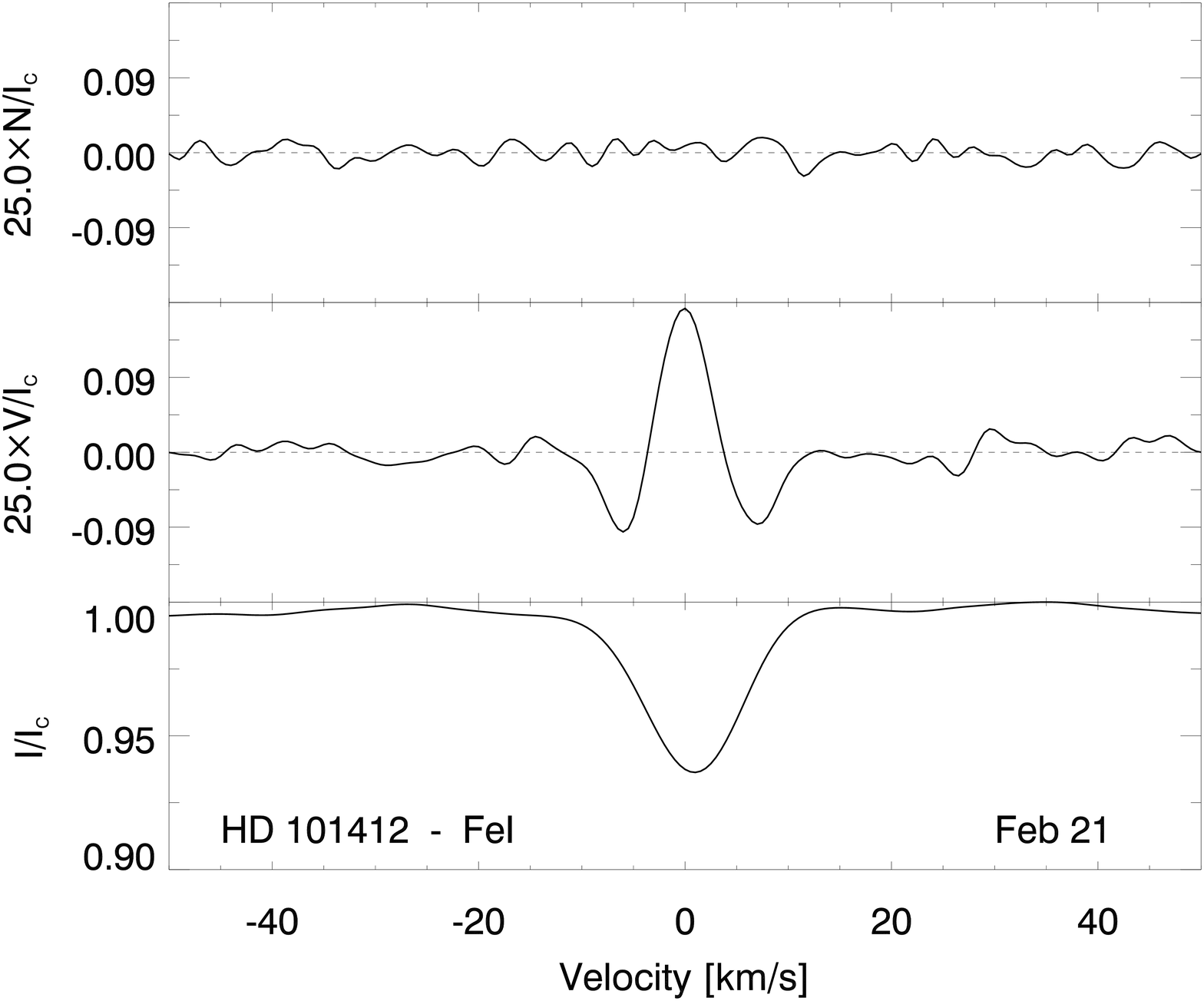}
\includegraphics[width=0.20\textwidth]{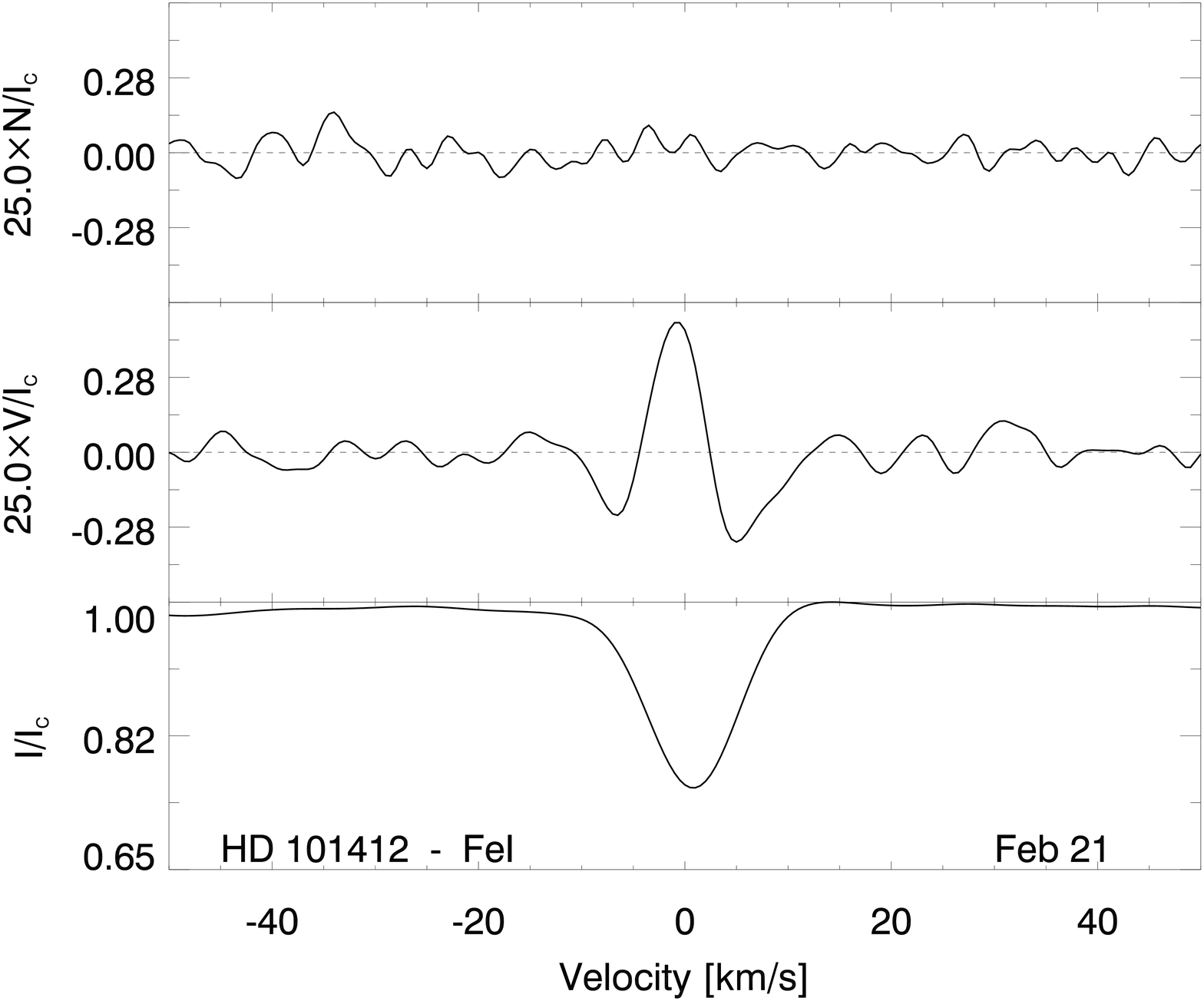}
\includegraphics[width=0.20\textwidth]{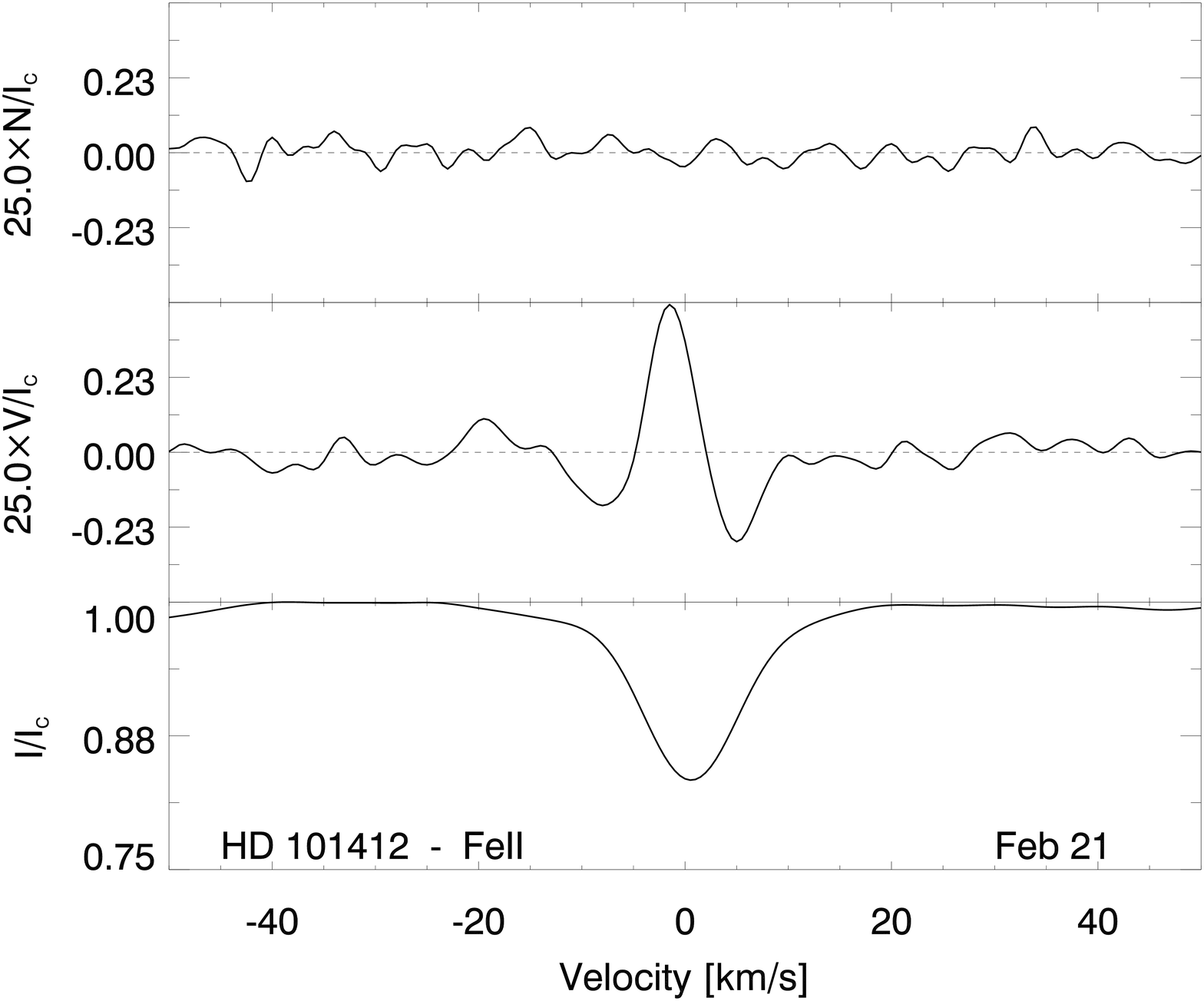}
\includegraphics[width=0.20\textwidth]{Fi2f.eps}
\includegraphics[width=0.20\textwidth]{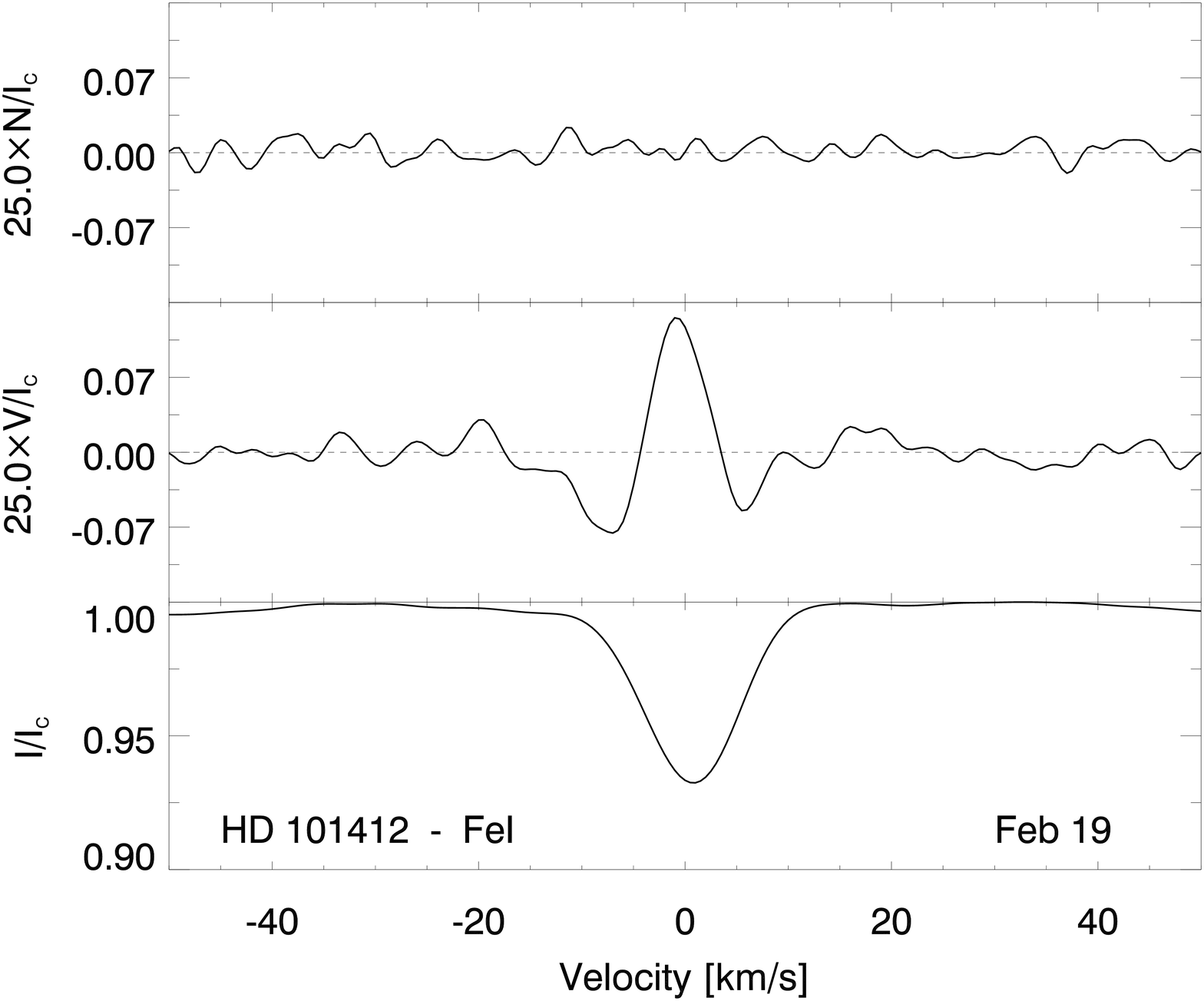}
\includegraphics[width=0.20\textwidth]{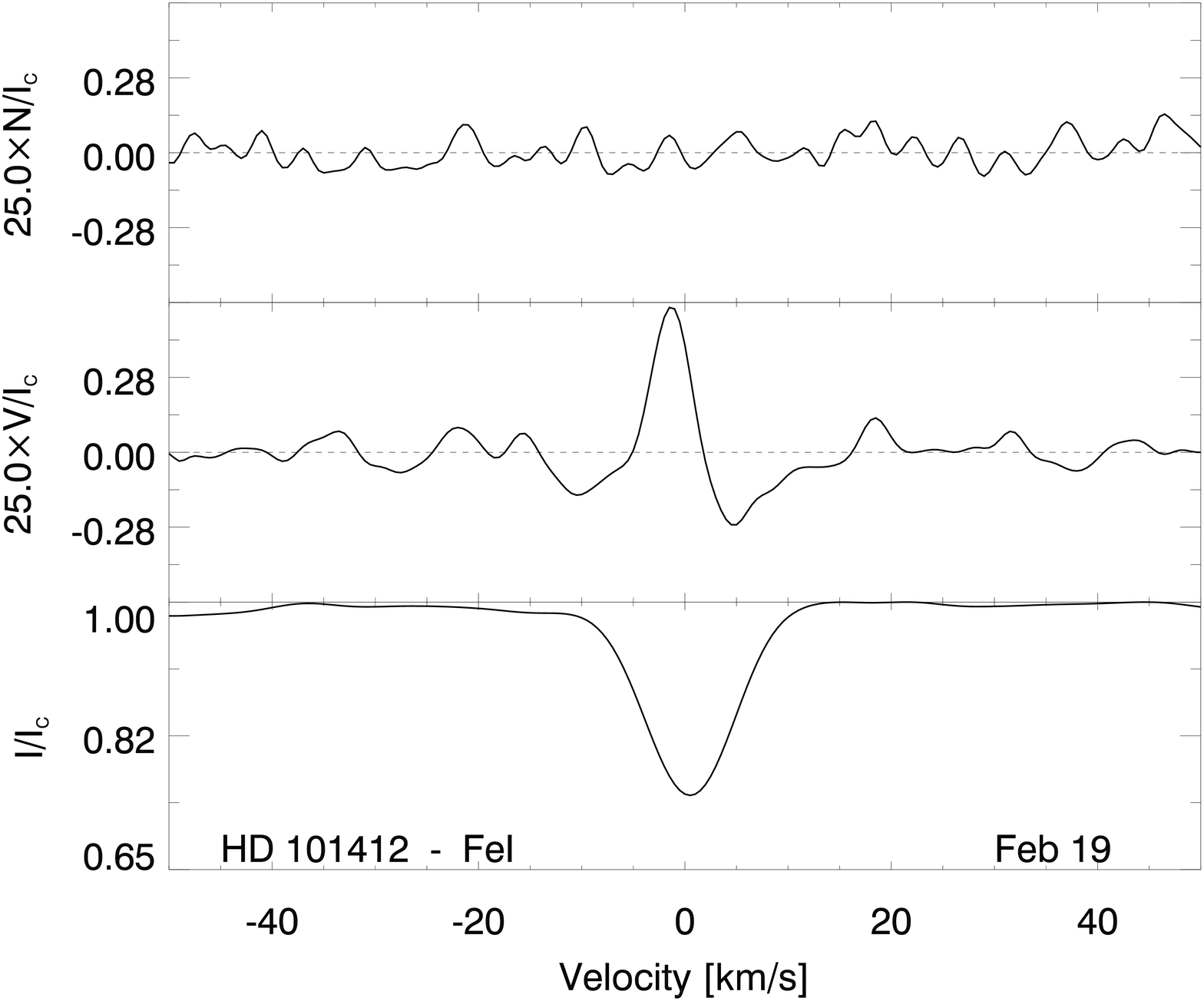}
\includegraphics[width=0.20\textwidth]{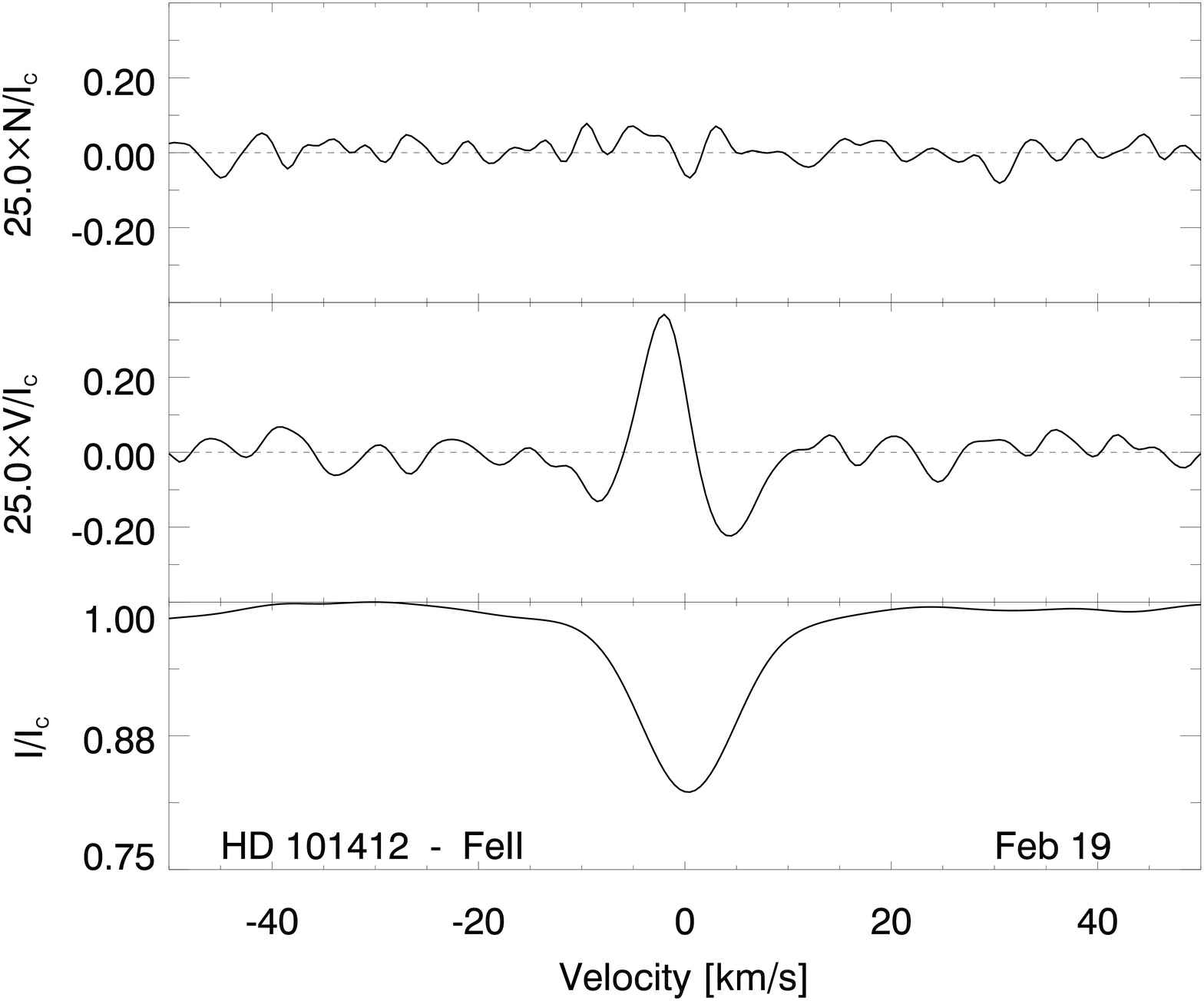}
\includegraphics[width=0.20\textwidth]{Fi2e.eps}
\includegraphics[width=0.20\textwidth]{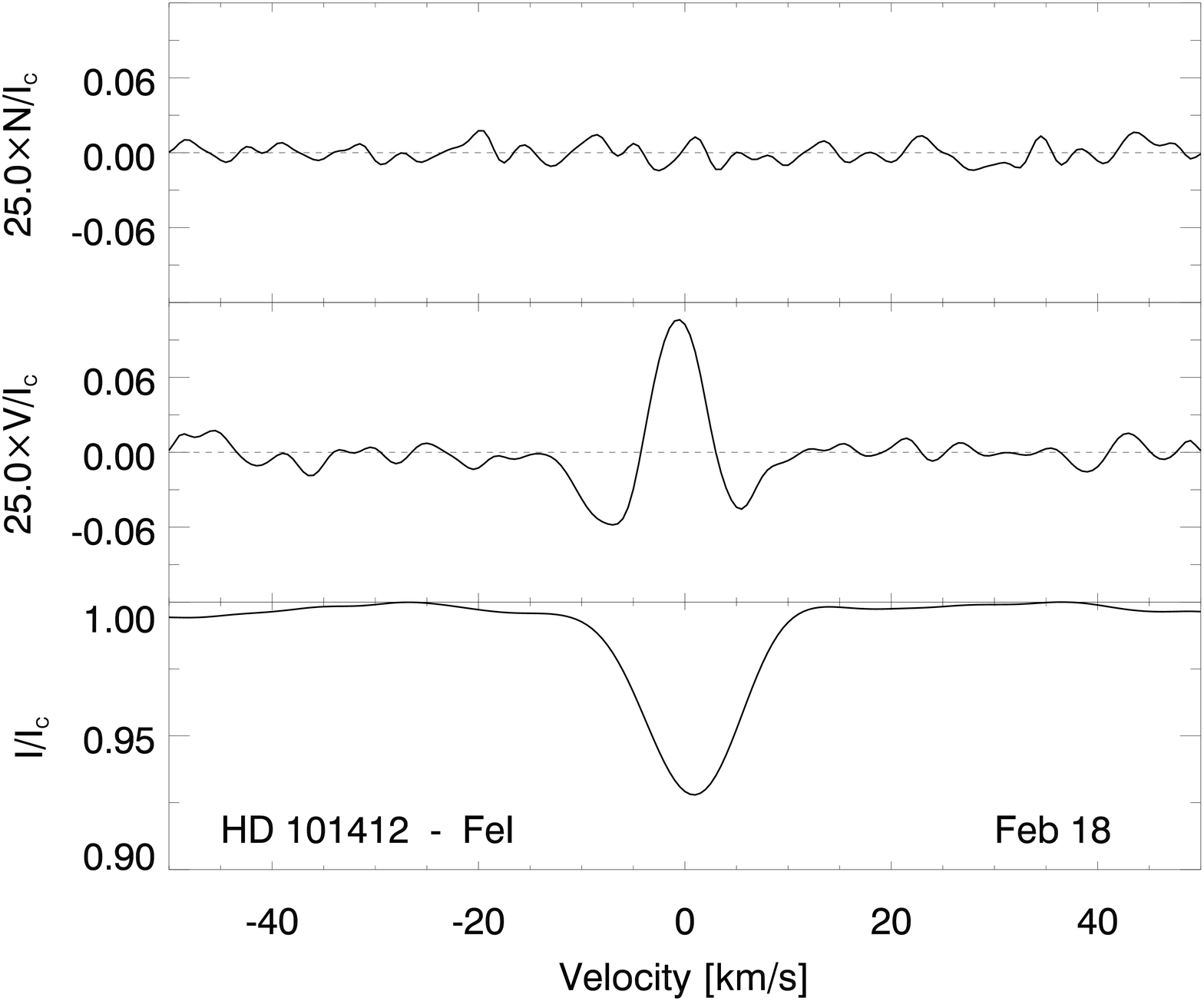}
\includegraphics[width=0.20\textwidth]{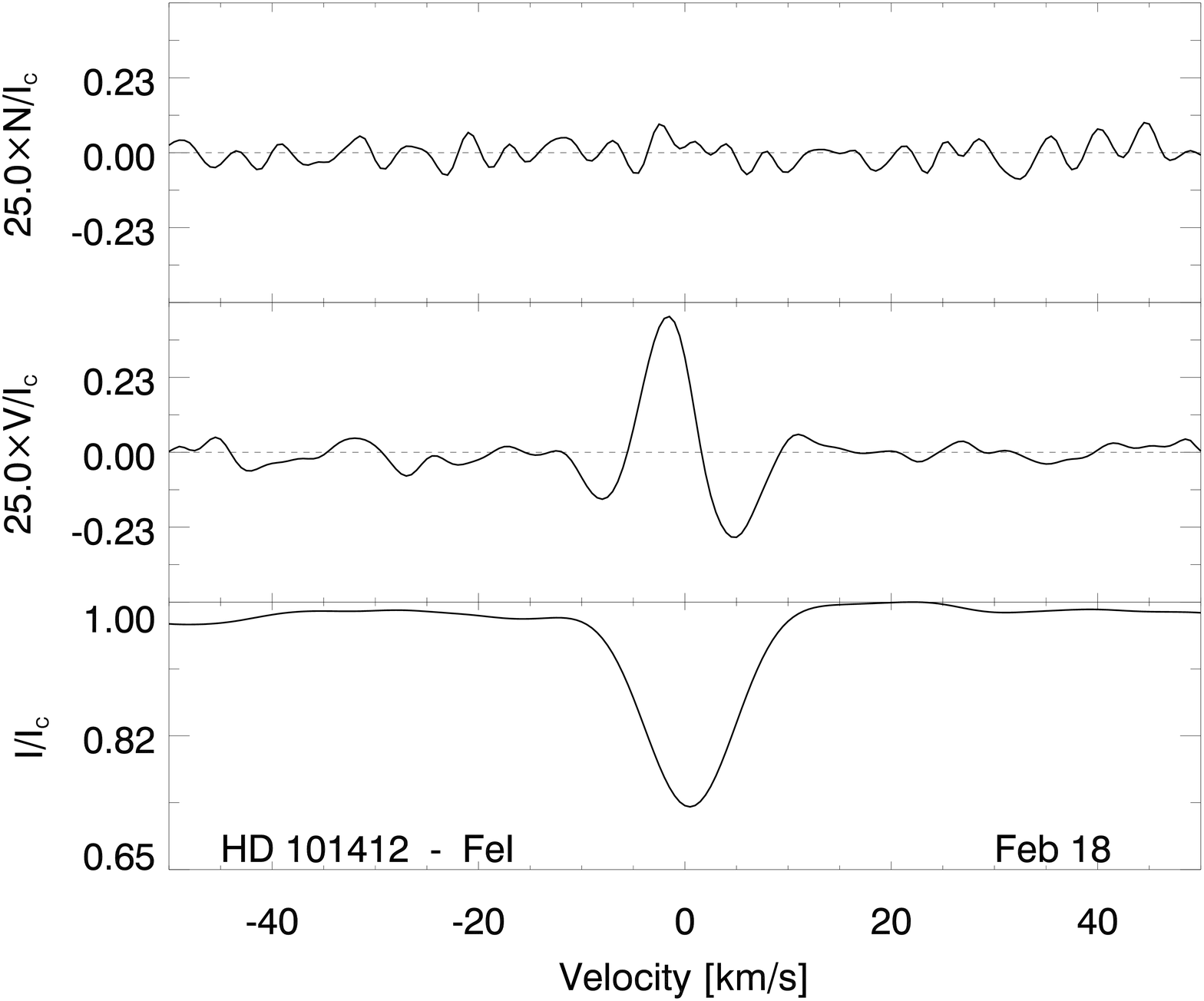}
\includegraphics[width=0.20\textwidth]{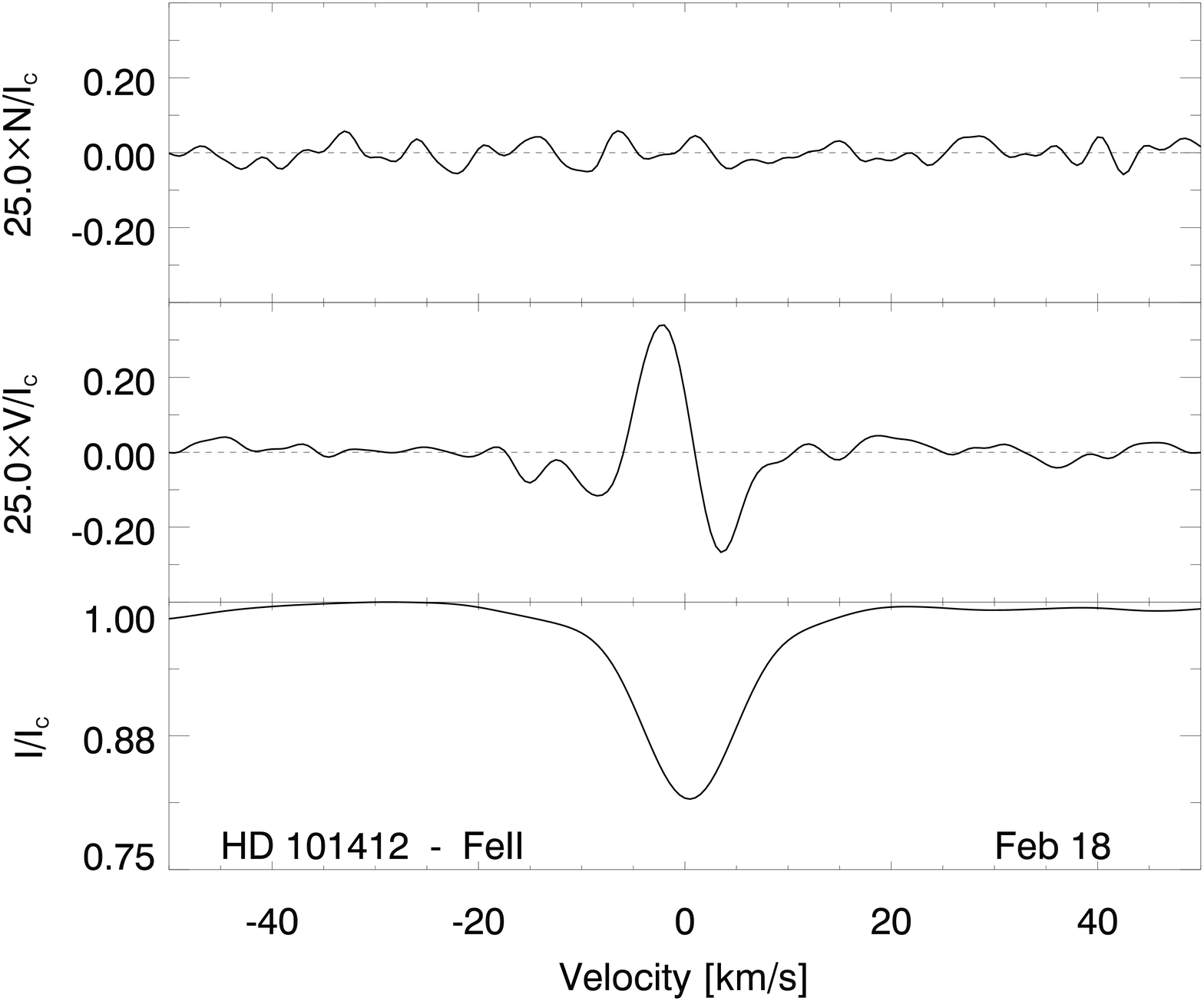}
\includegraphics[width=0.20\textwidth]{Fi2d.eps}
\includegraphics[width=0.20\textwidth]{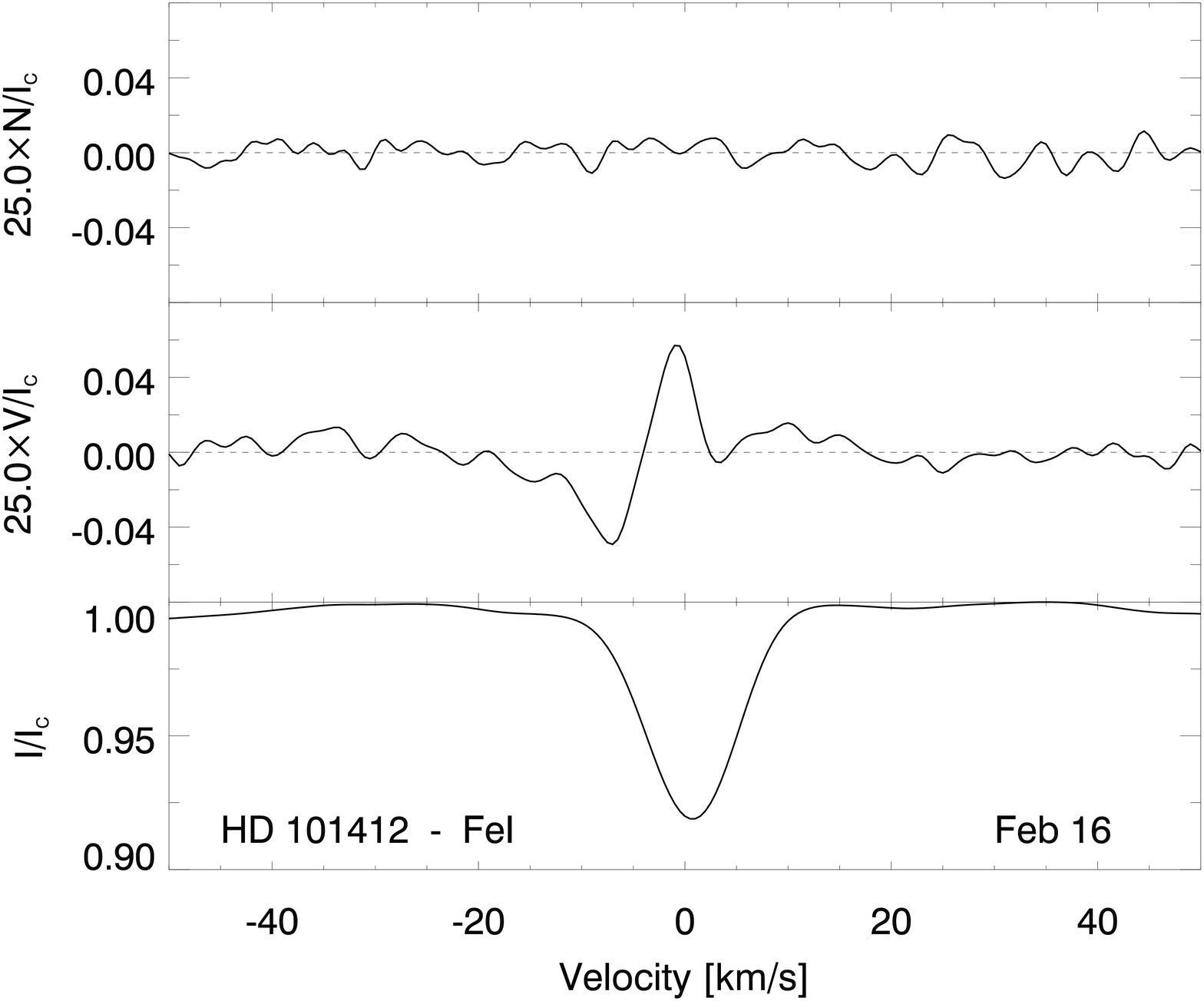}
\includegraphics[width=0.20\textwidth]{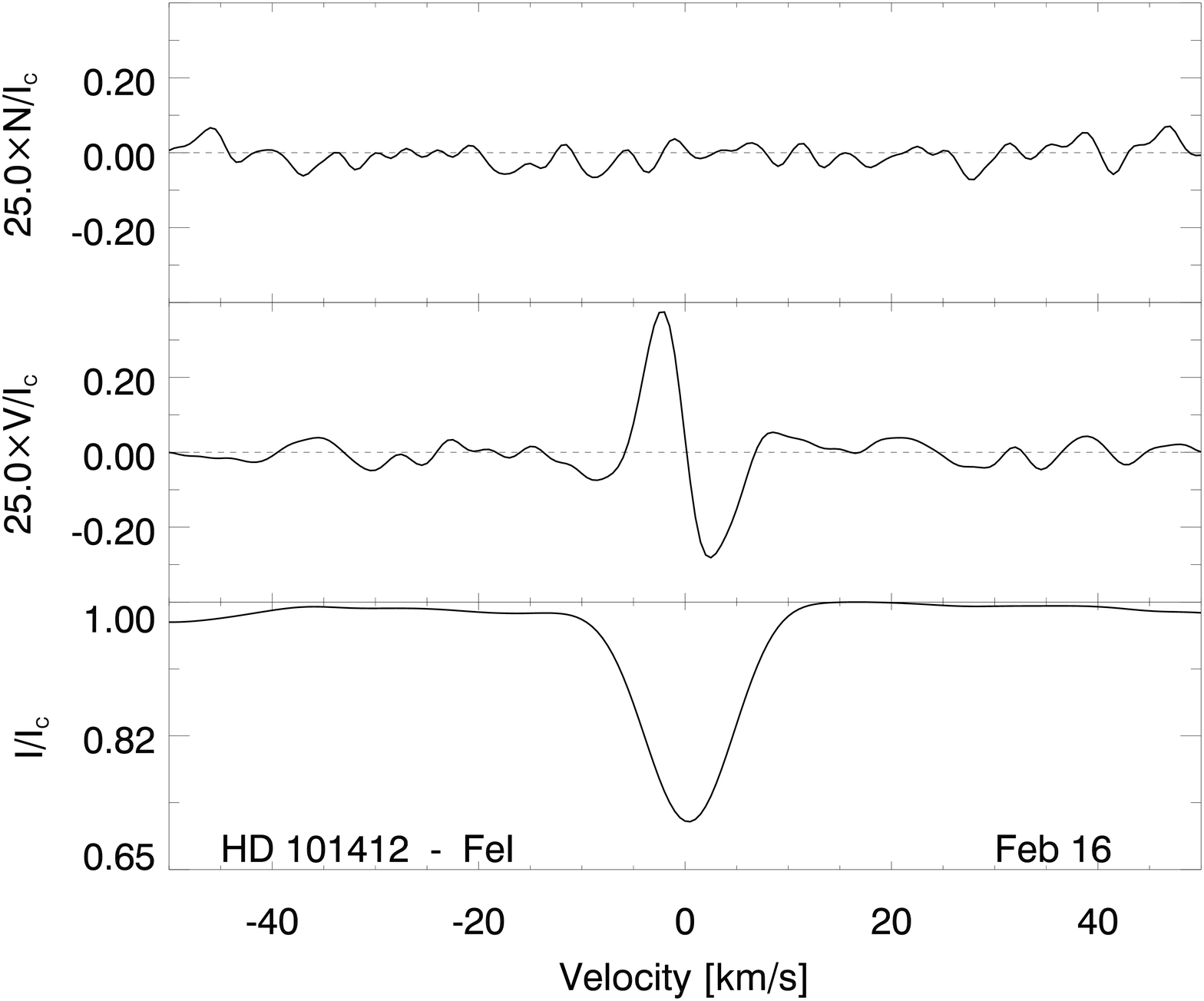}
\includegraphics[width=0.20\textwidth]{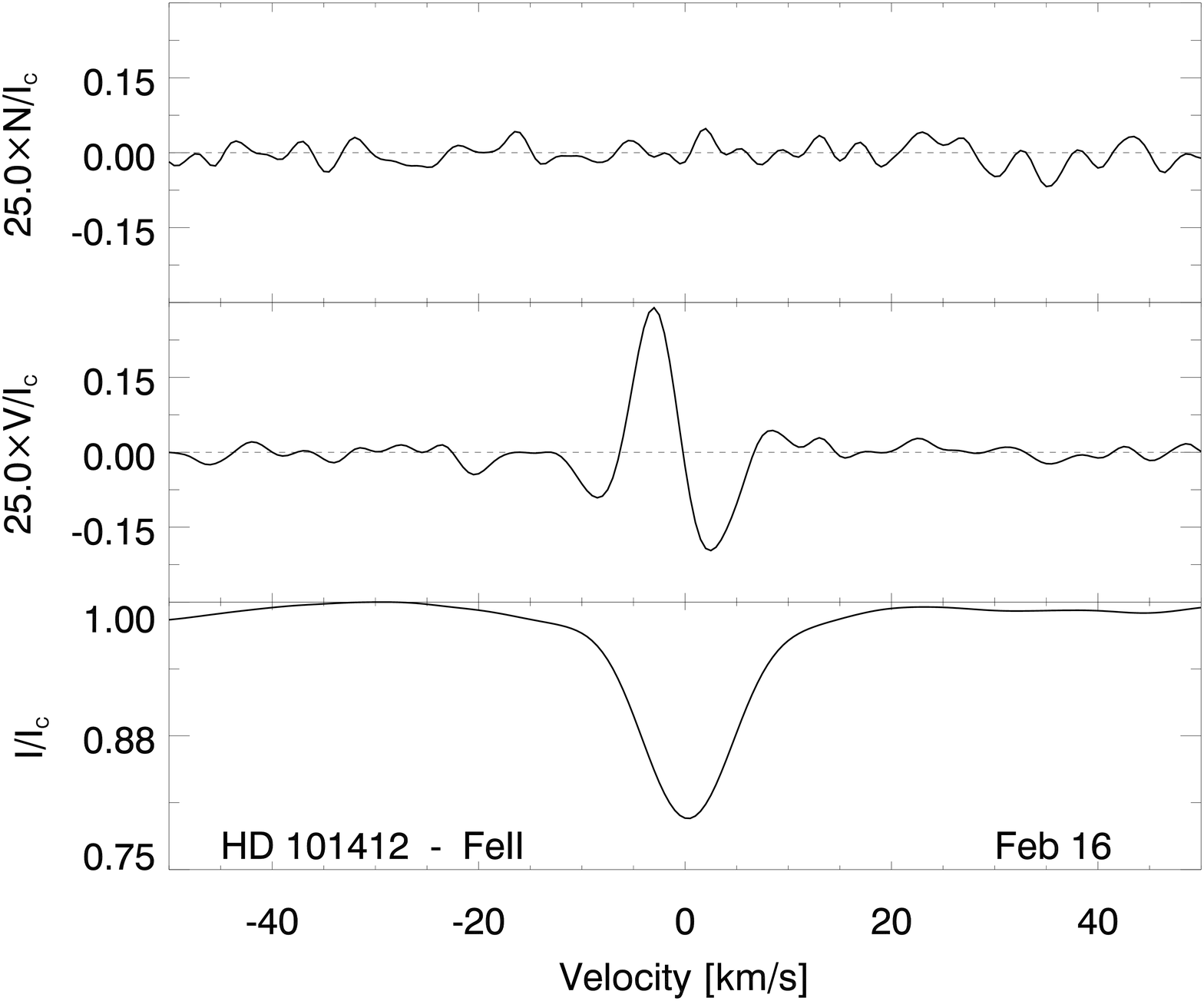}
\includegraphics[width=0.20\textwidth]{Fi2c.eps}
\includegraphics[width=0.20\textwidth]{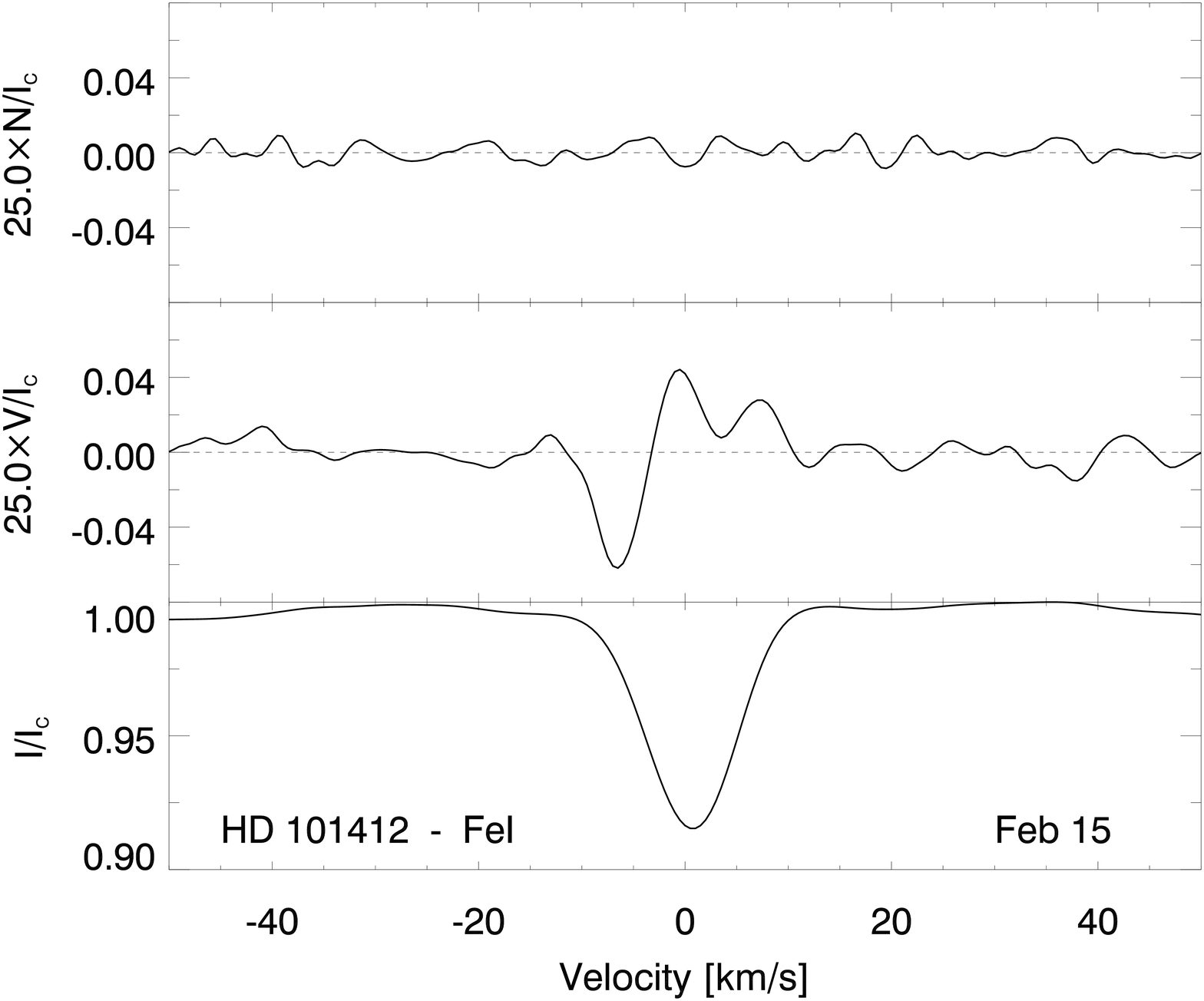}
\includegraphics[width=0.20\textwidth]{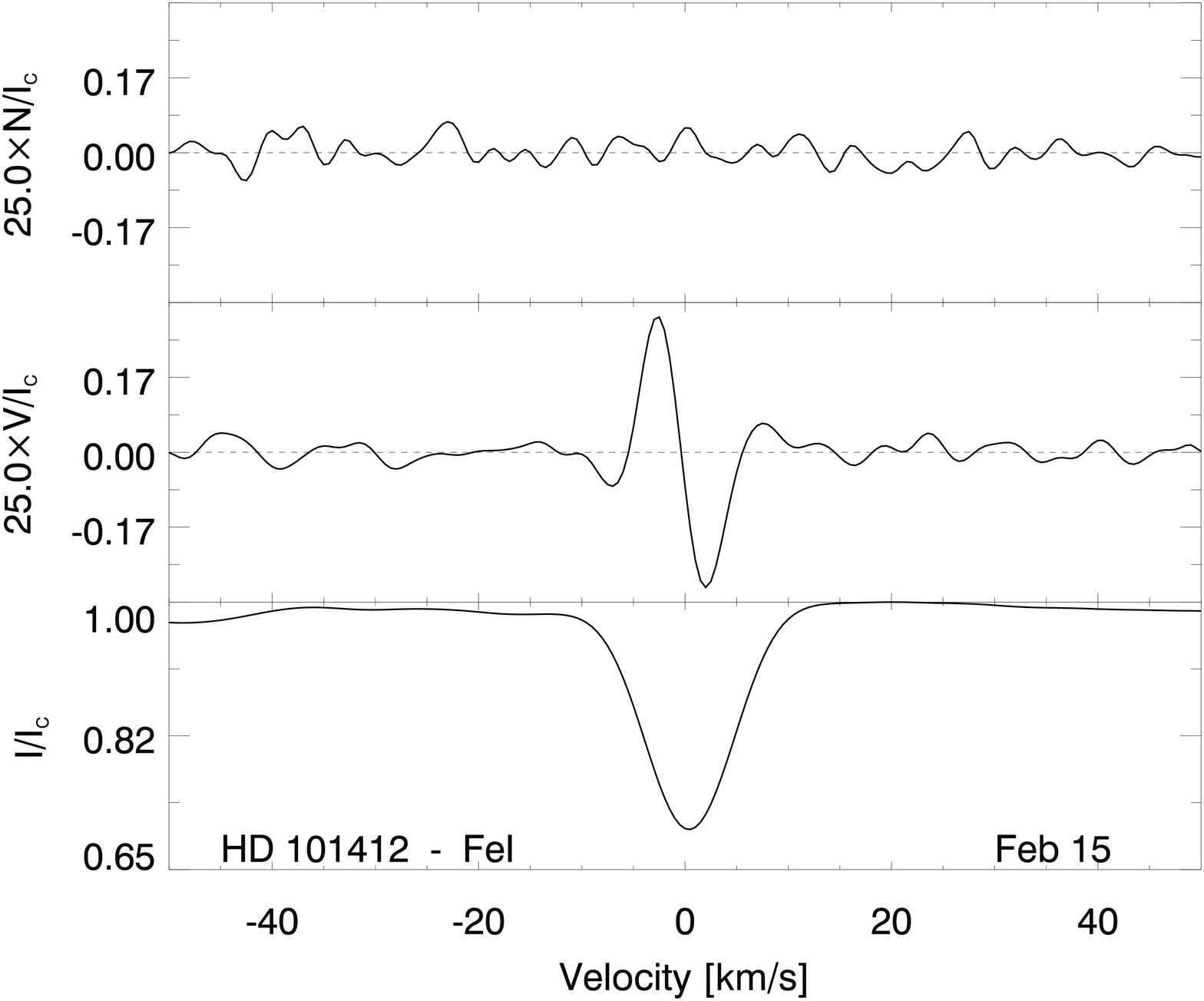}
\includegraphics[width=0.20\textwidth]{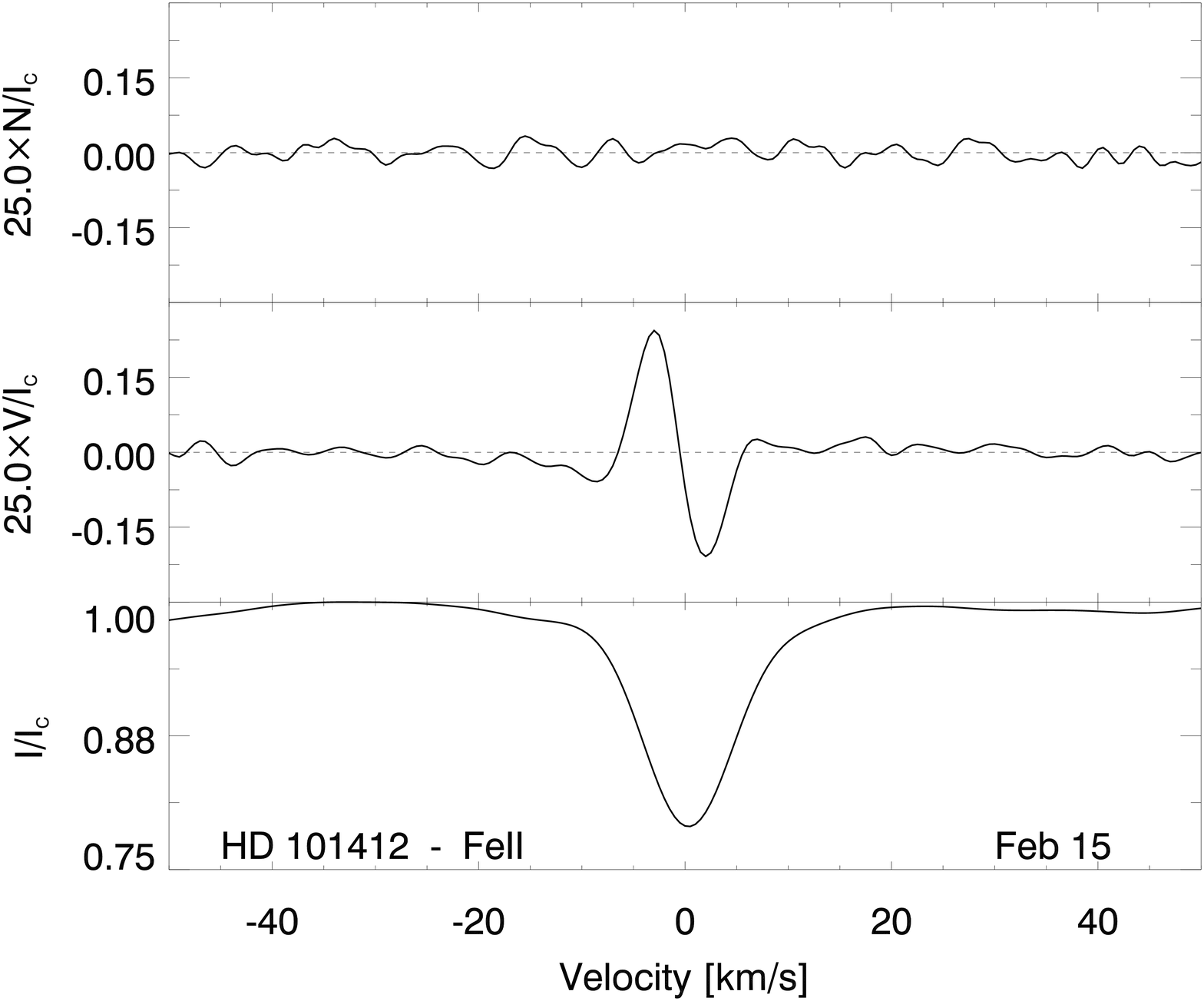}
\includegraphics[width=0.20\textwidth]{Fi2b.eps}
\includegraphics[width=0.20\textwidth]{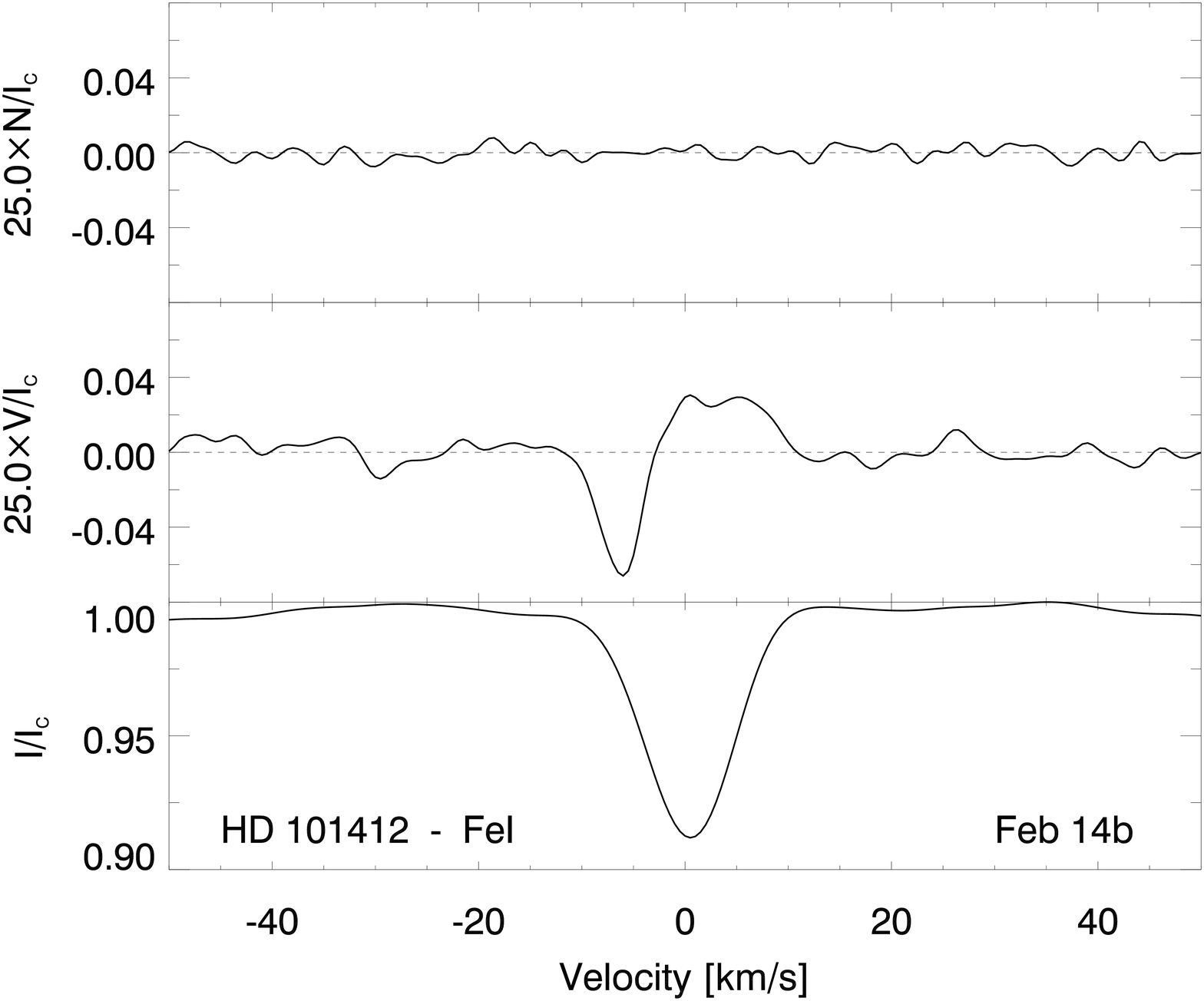}
\includegraphics[width=0.20\textwidth]{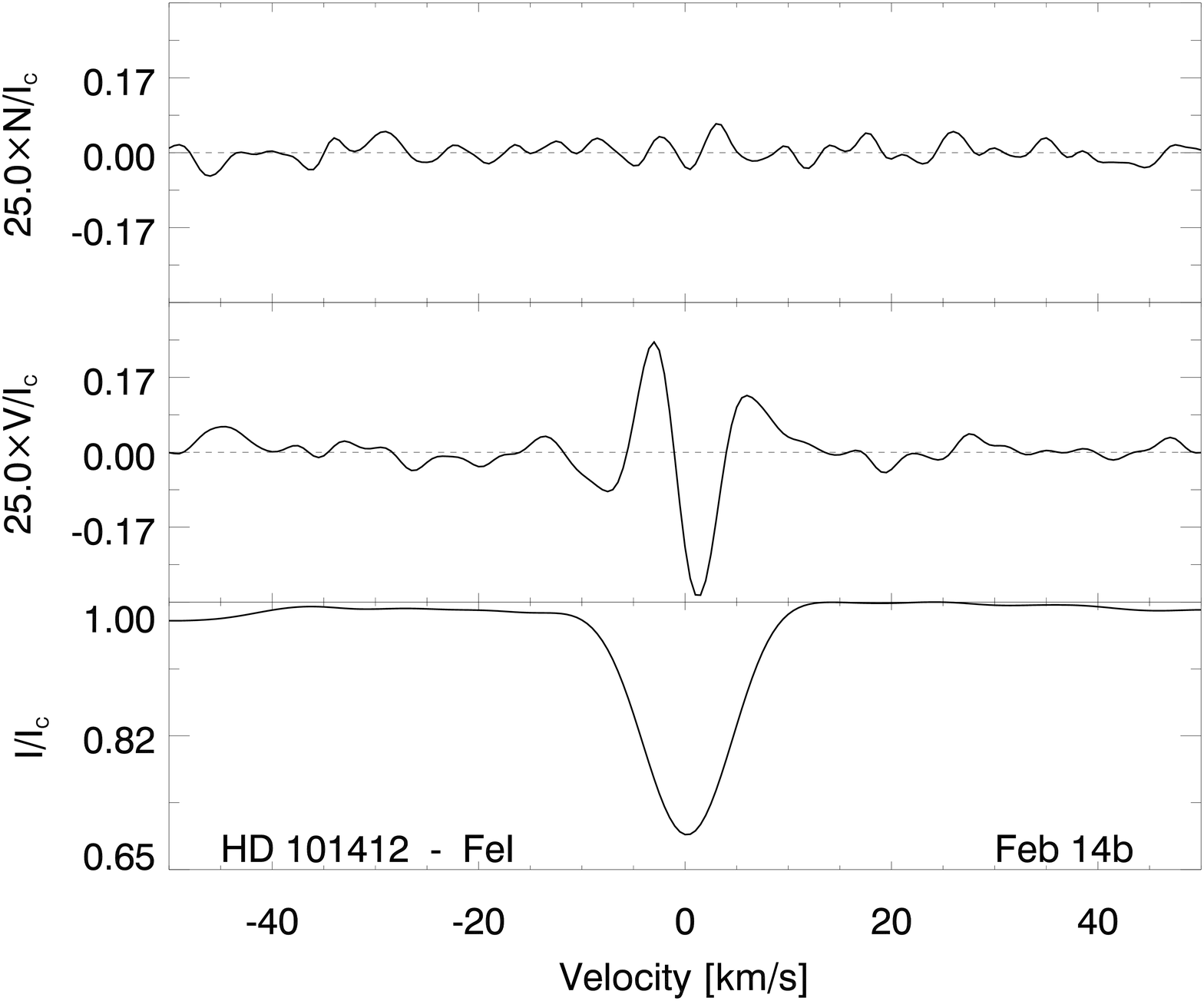}
\includegraphics[width=0.20\textwidth]{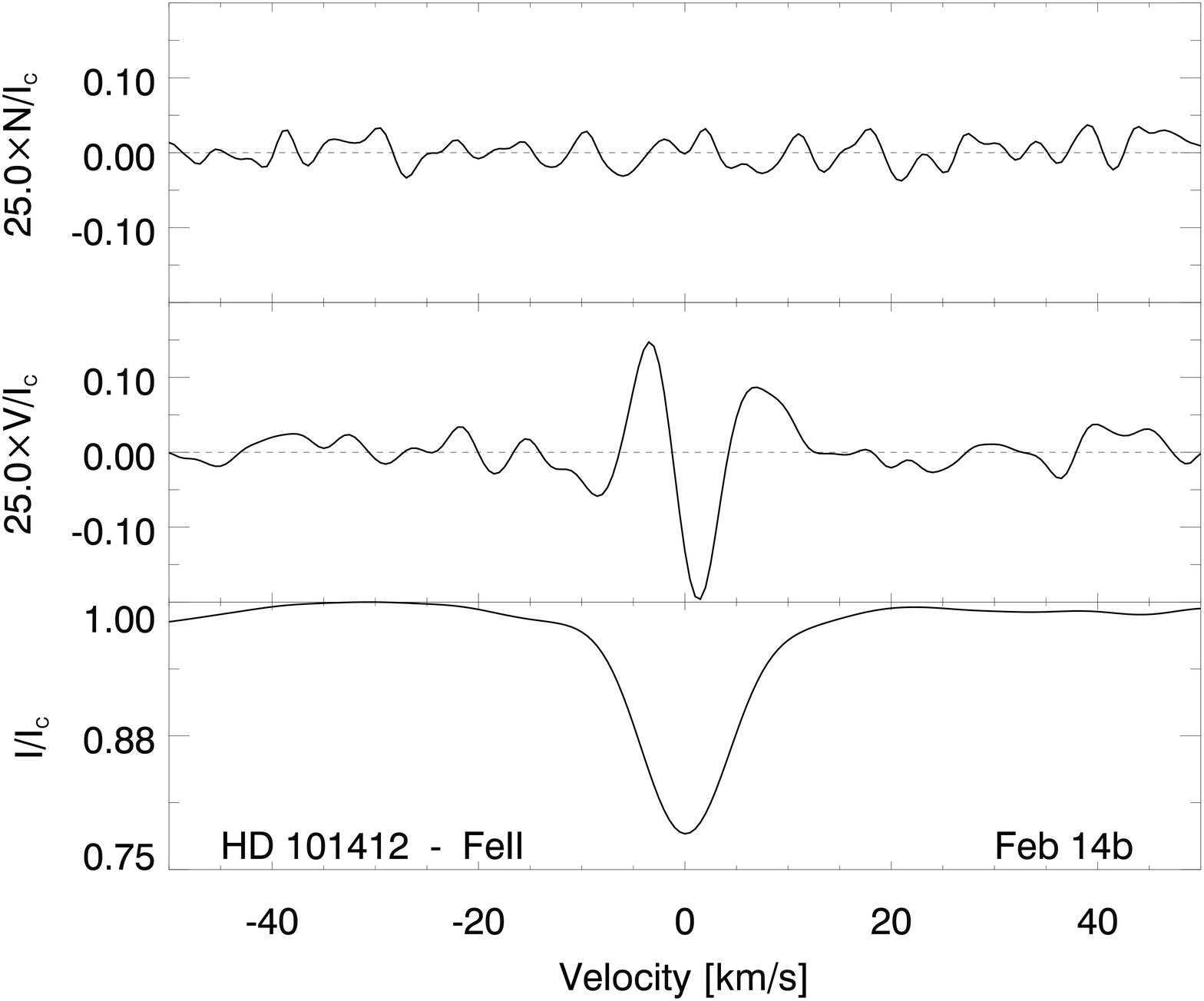}
\includegraphics[width=0.20\textwidth]{Fi2a.eps}
\includegraphics[width=0.20\textwidth]{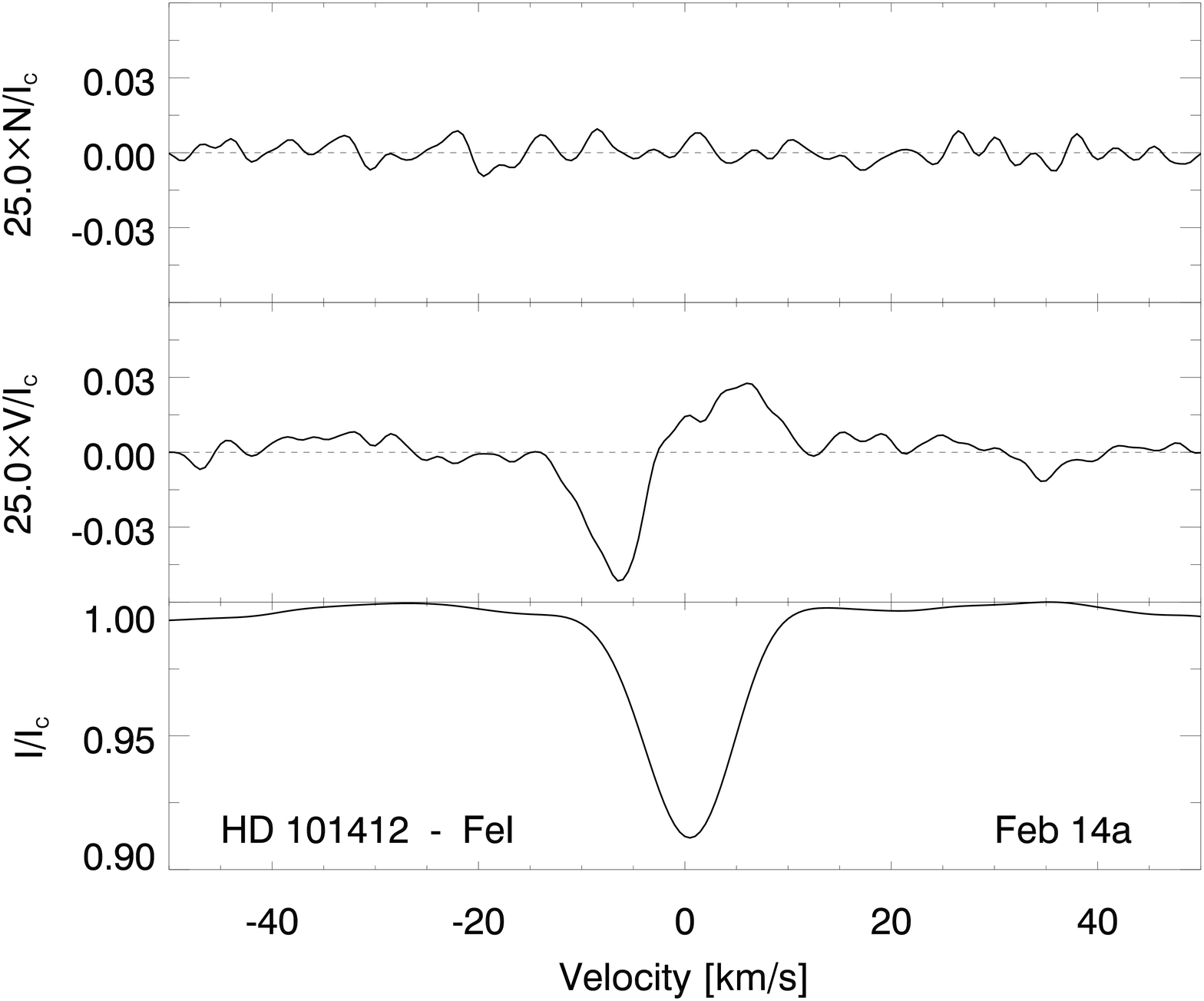}
\includegraphics[width=0.20\textwidth]{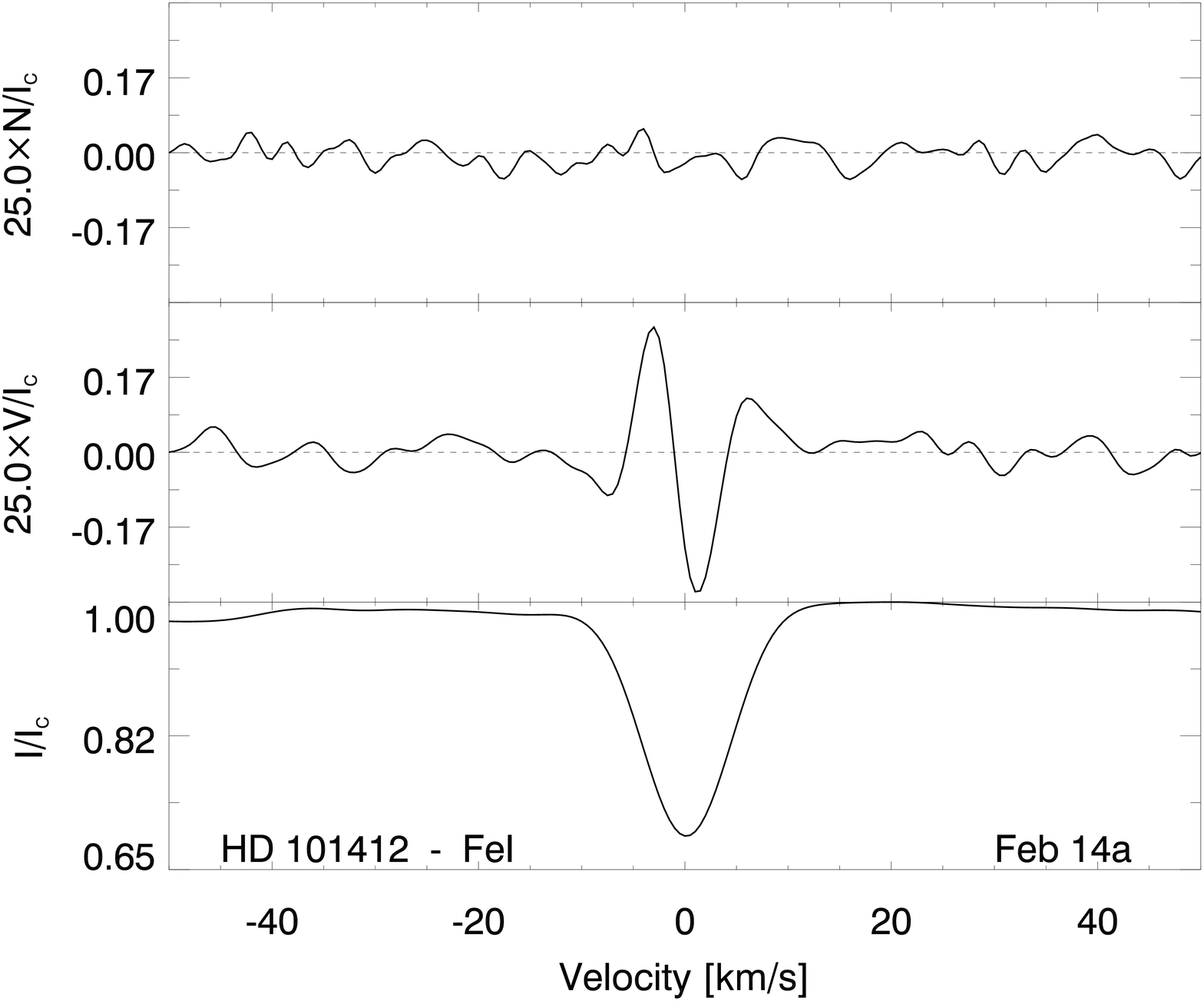}
\includegraphics[width=0.20\textwidth]{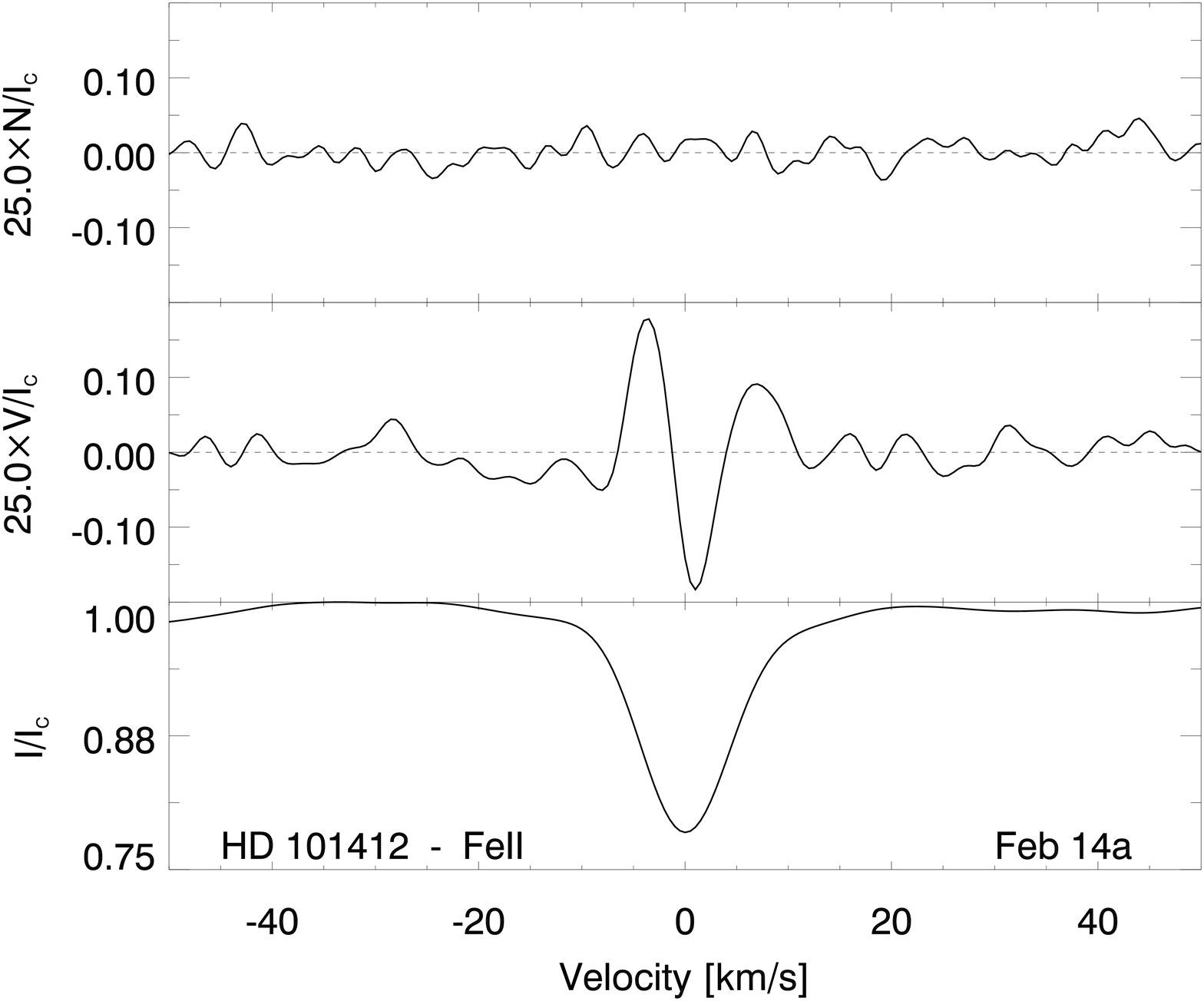}
\caption{ The correspondence of SVD $I$ and $V$ profiles of HD\,101412 
obtained using different line masks.
From left to right the results correspond to the following line samples:
the sample of 650 lines belonging to various iron-peak elements assuming 
$T_{\rm eff}=8\,300$\,K, 
the sample of 339 \ion{Fe}{i} lines ($T_{\rm eff}=8\,300$\,K), 
the sample of 29 \ion{Fe}{i} lines assuming $T_{\rm eff}=10\,000$\,K,
and the sample of 52 \ion{Fe}{ii} lines ($T_{\rm eff}=8\,300$\,K).
The time runs from bottom to top.} 
\label{fig:AppSVDall}
\end{figure*}

\end{document}